\documentclass[aps, prx, reprint, longbibliography, nofootinbib,superscriptaddress, floatfix]{revtex4-1}
\usepackage{slashed}
\usepackage{verbatim}
\usepackage[T1]{fontenc}
\usepackage{mathbbol}
\usepackage[dvipsnames]{xcolor}
\usepackage{orcidlink}

\usepackage[normalem]{ulem}
\usepackage[english]{babel}
\usepackage{lipsum}
\usepackage{physics}
\usepackage{dcolumn}
\usepackage{tensor}
\usepackage{comment}
\usepackage{xcolor}
\usepackage{graphicx,color,overpic,mathtools}
\usepackage{amsthm,amsmath,amssymb,mathrsfs}
\usepackage{braket,bm,bbm,setspace}
\usepackage{cancel, float, xargs}
\definecolor{myred}{RGB}{179, 27, 27}
\definecolor{Dmagenta}{RGB}{255, 0, 255}
\usepackage{hyperref}
\hypersetup{
    colorlinks=true,
    linkcolor=myred, 
    citecolor=myred, 
    urlcolor=myred  
 }
\usepackage{siunitx}
\newcommand{\mb}{m_{\rm b}}

\begin{document}

\raggedbottom

\title{Black Hole Ringdown Nonlinearities in the Large-$D$ Limit}

\author{Roberto Emparan\,\orcidlink{0000-0003-2457-0203}}
\email{emparan@ub.edu}
\affiliation{Institució Catalana de Recerca i Estudis Avançats (ICREA), Passeig Lluis Companys, 23, 08010
Barcelona, Spain}
\affiliation{Departament de Física Quàntica i Astrofísica and Institut de Ciències del Cosmos, Universitat de
Barcelona, Martí i Franquès, 1, 08028 Barcelona, Spain}

\author{Amanda Green-Salinas\,\orcidlink{0009-0005-1614-1144}}
\email{amanda.greensalinas@ub.edu}
\affiliation{Departament de Física Quàntica i Astrofísica and Institut de Ciències del Cosmos, Universitat de
Barcelona, Martí i Franquès, 1, 08028 Barcelona, Spain}

\author{David Pereñiguez\,\orcidlink{0000-0002-1007-4551}}
\email{dpereni1@jhu.edu}
\affiliation{William H. Miller III Department of Physics \& Astronomy, Johns Hopkins University\\
3400 North Charles Street, Baltimore, MD 21218, USA}

\author{Jaime Redondo-Yuste\,\orcidlink{0000-0003-3697-0319}}
\email[]{jredondo@jhu.edu}
\affiliation{William H. Miller III Department of Physics \& Astronomy, Johns Hopkins University\\
3400 North Charles Street, Baltimore, MD 21218, USA}

\begin{abstract}
We initiate the study of nonlinear effects in the ringdown phase of black hole mergers using the effective theory of black hole dynamics in the large-$D$ limit. This framework offers several advantages: the quasinormal mode spectrum, including nonlinear corrections, is analytically tractable; numerical simulations of collisions are computationally inexpensive; and the extraction and analysis of the ringdown signal are clean and controlled. As a proof of concept, we derive analytic expressions for the third-order response of a static black hole driven by a single quasinormal mode, and apply them to study the ringdown following head-on collisions of non-spinning black holes across a range of velocities and mass ratios. We find that including nonlinear effects, up to quadratic and cubic order, improves the accuracy of quasinormal-mode modelling of black hole relaxation by several orders of magnitude. The results also show a clear growth in the strength of nonlinear effects as the collision velocity increases.
\end{abstract}

\maketitle
%\onecolumngrid
%\tableofcontents

%%%%%%%%%%%%%%%%%%%%%%
\section{Introduction}
%%%%%%%%%%%%%%%%%%%%%%

Gravitational wave observations of binary black hole mergers are entering a precision era in which subdominant features of the signal are both detectable and physically consequential. The ringdown—the final stage, when the remnant relaxes to a stationary black hole—is especially informative: it lies within perturbation theory, making a systematic treatment possible. While late-time signals are well described by linear quasinormal modes (QNMs), increasing detector sensitivity makes a quantitative understanding of nonlinear dynamics urgent. Numerical simulations already reveal nonlinear effects---higher harmonics---in the post-merger relaxation~\cite{London:2014cma, Cheung:2022rbm, Mitman:2022qdl} (see also~\cite{Zlochower:2003yh}), challenging the long-held view that this phase is dominated by linear fluctuations. Next-generation detectors are expected to detect such effects~\cite{Yi:2024elj,Lagos:2024ekd}, and current observations offer tantalizing hints of nonlinear signatures~\cite{Wang:2026rev}. 

Despite significant progress over the past three decades, key questions concerning nonlinear dynamics during ringdown remain open~\cite{Gleiser:1995gx, Gleiser:1996yc, Garat:1999vr, Campanelli:1998jv, Brizuela:2006ne, Brizuela:2007zza, Brizuela:2009qd, Pazos:2010xf, Ioka:2007ak, Nakano:2007cj, Lagos:2022otp,Bucciotti:2023ets,Khera:2023oyf,Bucciotti:2024zyp,Bucciotti:2024jrv,BenAchour:2024skv,Ma:2024qcv,Bourg:2024jme,Singh:2025xzd,Redondo-Yuste:2023seq, Khera:2024bjs, Spiers:2023cip, Spiers:2023mor, Bourg:2025lpd}. These include, for example, clarifying the role of pseudo-resonances (a third-order effect)~\cite{Yang:2014tla, Sberna:2021eui, Redondo-Yuste:2023ipg, Zhu:2024dyl, May:2024rrg}, and establishing the domain of convergence of black hole perturbation theory~\cite{LISAConsortiumWaveformWorkingGroup:2023arg}. 
The intrinsic complexity of the problem has kept these issues largely unresolved. Any insight --- even at a qualitative level --- into the importance of nonlinearities during ringdown, and their dependence on progenitor parameters would therefore be valuable. This motivates the use of simplified frameworks that retain the nonlinear structure of Einstein’s equations while remaining tractable enough to yield clear physical intuition.

In this work we study ringdown beyond linear order using the effective theory of black hole dynamics in the large-dimension limit, $D\to\infty$ \cite{Emparan:2013moa,Emparan:2015hwa,Bhattacharyya:2015dva,Bhattacharyya:2015fdk,Emparan:2020inr,Emparan:2025yfy}. In this limit the problem simplifies dramatically for two reasons: the gravitational field is strongly localized near the horizon, and gravitational radiation couples to the black hole only non-perturbatively in $1/D$. As a result, near-horizon dynamics decouples from far-zone emission and is captured by effective equations that encode the full nonlinear structure of Einstein’s equations but are far simpler to solve \cite{Emparan:2015gva}. We will show that with this framework we can analyze in detail how the horizon forms and relaxes during a merger, accounting for nonlinearities in a simple, explicit manner. Although the asymptotic %\dpr{(I'd add this word because technically what we look at are also GWs, although those associated to horizons)}\RE{OK to add asymptotic. I'm not sure I'd say we're looking at GWs associated to horizons, since there's no radiation zone; but they're oscillations of the geometry i.e., of the gravitational field, and at this looser level it's a matter of semantics to call them GWs -- so OK} 
gravitational-wave signal---absent in our analysis---can in principle be recovered by reintroducing the coupling to the far zone \cite{Bhattacharyya:2016nhn}, this is not needed to characterize the size and properties of nonlinear effects in ringdown, which is our primary goal here.

A crucial advantage of the large-$D$ limit is that the linear quasinormal mode spectrum of static and rotating black holes admits a closed analytic form \cite{Emparan:2014aba,Emparan:2015rva}, and we will show that this analytic control extends to arbitrary nonlinear orders in perturbation theory. We should be upfront, however, that the spectra of the Schwarzschild black hole and its large-$D$ counterpart differ in important ways: axial and polar perturbations are no longer isospectral for $D\geq 5$, damping is typically stronger, and the wider mode spacing makes nonlinear quasiresonances harder to realize. Nevertheless, the large-$D$ effective theory has repeatedly provided accurate qualitative guidance---and even predictions---for highly nontrivial black hole dynamics at finite $D$ \cite{Emparan:2018bmi,Andrade:2019edf,Andrade:2020dgc,Emparan:2023dxm}. We expect that, when interpreted with care, the lessons extracted here will likewise offer useful insight into nonlinear ringdown.

\begin{figure}%[t]
    \centering
    \includegraphics[width=\columnwidth]{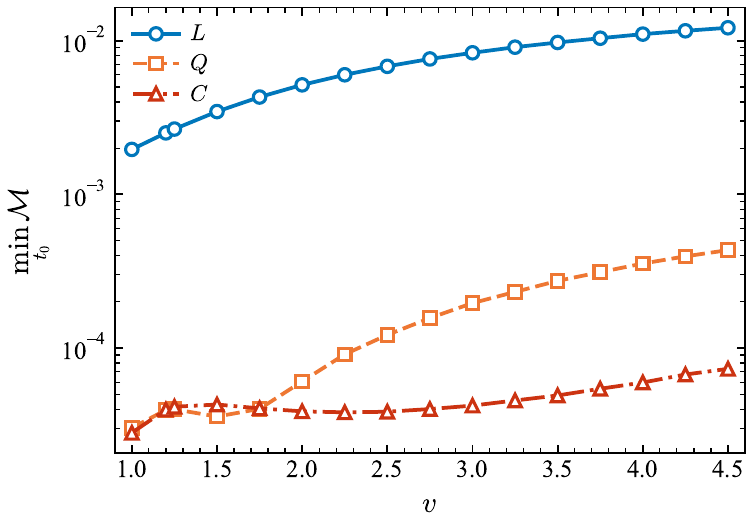}
    \caption{Minimum mismatch achieved by ringdown models including only linear QNMs (L), linear plus quadratic effects (Q), and linear plus quadratic and cubic effects (C), as a function of the impact velocity $v$ in a head-on black-hole collision. All three models contain the same number of free parameters (two). The nonlinear models improve the fit by more than one order of magnitude, with the cubic model outperforming the quadratic one at high merger velocities. }
    \label{fig:mismatch}
\end{figure}

In this article we initiate this program in the simpler setting of non-rotating black holes. We study fluctuations of a static black hole and present an algorithmic procedure to construct nonlinear quasinormal mode solutions at arbitrary order in perturbation theory. We then apply this to derive analytical expressions for the third-order response driven by a single quasinormal mode. This allows us to: (i) confirm that nonlinear effects are robustly present---at least through cubic order---in the ringdown following a head-on collision of two non-spinning black holes, across a range of mass ratios and collision velocities (see Fig.~\ref{fig:mismatch}), and (ii) show in a transparent way that neglecting nonlinearities can lead to biased estimates of the ringdown content (see Fig.~\ref{fig:evidence}). 

The article is organised as follows. In Section~\ref{sec:BHDyn}, we review the effective description of black hole dynamics in the large-$D$ limit. In Section~\ref{sec:BHPT}, we discuss black hole perturbation theory within this framework and construct nonlinear quasinormal mode solutions. In Section~\ref{sec:Head_On_Black_Hole_Mergers}, we present numerical simulations of head-on mergers and analyse the ringdown, finding clear evidence of nonlinear effects. We conclude in Section~\ref{sec:Discussion}, where we summarise our results and outline directions for future work.

%%%%%%%%%%%%%%%%%%%%%%%%%%%%%%%%%%%%%%
\section{Black Hole Dynamics in the Large-$D$ Limit}\label{sec:BHDyn}
%%%%%%%%%%%%%%%%%%%%%%%%%%%%%%%%%%%%%%

The large-$D$ effective theories of \cite{Emparan:2015gva,Emparan:2016sjk} describe metrics of the Eddington-Finkelstein form
\begin{align}
    ds^2 = & 2 dt\,dr-\left( 1- \frac{m(t,z)}{\sf R}\right)dt^2 -\frac{2}{D}\frac{p(t,z)}{\sf R}dt\,dz \nonumber \\ &+ \frac1{D}\left(1+\frac1{D}\frac{p^2(t,z)}{ m(t,z)} \frac{1}{\sf R}\right)dz^2+ r^2 d\Omega_{D-3}\, \label{eq:EFmetric}
\end{align}
where the coordinate
\begin{equation}
    {\sf R}=r^{D-3}
\end{equation}
remains finite as $D\to\infty$. These metrics describe black holes and black strings with horizons at ${\sf R} =m(t,z)$\footnote{To leading order in $1/D$, the apparent and event horizons coincide.} extending along the direction $z$ and the homogeneous sphere $S^{D-3}$.\footnote{More generally, the horizon can vary along any finite number of directions $z^i$, $i=1,\dots,p$, times a homogeneous $S^{D-p-2}$, but in this article we only need one $z$.} The factors of $D$ in them are chosen so that, when this ansatz is substituted into the Einstein equations, all the functions appearing in \eqref{eq:EFmetric} enter at the same order in $1/D$. The radial dependence in $r$ (or ${\sf R}$) has already been solved explicitly, and the remaining Einstein equations are satisfied whenever $m(t,z)$ and $p(t,z)$ are solutions of
\begin{equation}\label{eq:effective_eqs}
    \begin{aligned}
        \partial_t m - \partial^2_z m =& -\partial_z p \, , \\
        \partial_t p - \partial^2_z p =& \partial_z m -\partial_z\Bigl(\frac{p^2}{m}\Bigr) \, .
    \end{aligned}
\end{equation}
These are the effective equations for a dynamical black hole---directly encoding Einstein's equations in the limit $D\to\infty$. They are linear except for the last term $\partial_z(p^2/m)$, which makes nonlinear effects far more tractable analytically than in the full theory.

To see concretely how black holes are described in this framework, consider the
$D$-dimensional Schwarzschild-Tangherlini black hole,
\begin{align}
        ds^2=&-\left(1-\frac{m_0}{\rho^{D-3}}\right) d{\hat t}^2 + \frac{d\rho^2}{1-\frac{m_0}{\rho^{D-3}}} \notag\\
        &+\rho^2 \left(d\theta^2 + \cos^2\theta\, d\Omega_{D-3}\right)\label{eq:SchD}
\end{align}
with constant $m_0$ and take its large-$D$ limit via the coordinate change $(\rho,\theta)\to (r,z)$,
\begin{equation}
    r=\rho\, \cos\theta\,,\qquad z=\sqrt{D}\,\rho\, \sin\theta\,,
\end{equation}
keeping $z$ and $r^{D-3}={\sf R}$ finite as $D\to\infty$. Since
\begin{equation}
    \rho^2=r^2 +\frac{z^2}{D}\,,
\end{equation}
we have, as $D\to\infty$,
\begin{equation}
    \rho^{D-3}={\sf R}\left(1+ \frac{z^2}{D {\sf R}^{\frac2{D-3}}}\right)^{\frac{D-3}2}\to {\sf R}\, e^{z^2/2}\,.
\end{equation}
After this change of variables and redefining 
\begin{equation}
    t={\hat t} +\frac1{D}\ln \left({\sf R}- m_0\, e^{-z^2/2}\right)\,,
\end{equation}
the metric takes the form \eqref{eq:EFmetric} \cite{Andrade:2018nsz}, with
\begin{equation}
    m(z)=m_0\, e^{-z^2/2}\,, \quad p(z)=\partial_z m(z) =-m_0\, z\, e^{-z^2/2}\,.\label{eq:gblob}
\end{equation}
The key point is that, rather than starting from a known finite-$D$ solution and then taking the limit $D\to\infty$ as we have done, one can work directly with the effective equations~\eqref{eq:effective_eqs}, solve them to obtain \eqref{eq:gblob}, and then reconstruct the metric via \eqref{eq:EFmetric}. Given the simplicity of the effective equations \eqref{eq:effective_eqs}, this approach extends readily to problems that are far harder at finite $D$ than finding a static black hole solution---for instance, nonlinear perturbations of black holes or the collision of two black holes.\footnote{Rotating black holes can be incorporated by introducing at least one additional spatial direction, corresponding to a rotation plane $(z^1,z^2)$.}

In this approach, a spherical black hole is represented by a Gaussian blob profile such as \eqref{eq:gblob}. This reflects the fact that $m(z)$ encodes not the horizon radius itself, but its area-density along the polar direction $\theta$,  which in the large-$D$ limit becomes exponentially localized near the equator, $\theta \sim O(z/\sqrt{D})$, as follows from
\begin{equation}
    (\cos\theta)^{D-3}\simeq \left(1-\frac{z^2}{2D}\right)^{D-3}\to e^{-z^2/2}\,.
\end{equation}
As a result, the Gaussian profile extends over all $z$ rather than terminating at a finite distance. However, the large-$D$ approximation breaks down once $m \lesssim e^{-D}$, and we therefore regard two Gaussian blobs as effectively describing separate black holes whenever the density between them is sufficiently small. Strictly speaking, at leading order in large $D$ they remain connected by an extremely thin black bridge, but this subtlety plays no role for our purposes, since we focus on the evolution after the two blobs have merged into a single configuration.

The effective theory offers two further practical advantages beyond its simple nonlinear structure:
\begin{itemize}
    \item The equations \eqref{eq:effective_eqs} are of diffusion type, leading to a well-posed parabolic initial value problem. Their dissipative character also ensures very stable numerical behavior.

    \item 
    Gauge freedom and constraint issues are fully resolved at the level of the effective equations. As a result, initial data $m(0,z)$ and $p(0,z)$ can be specified freely and evolved without additional consistency conditions.
\end{itemize}
We now describe how head-on collisions are implemented in this framework. The equations admit a scaling symmetry that allows the overall amplitude to be fixed arbitrarily, as well as a Galilean invariance,
\begin{equation}
    z\to z-v t\,,\quad p\to p+ m v\,,
\end{equation}
with constant $v$.
A black hole moving at constant velocity $v$ along $z$ is therefore described by
\begin{equation}\label{eq:movingblob}
    m = m_0\,e^{-(z-vt)^2/2}\,, \quad p= -m_0\, (z-vt-v)\, e^{-(z-vt)^2/2}\,.
\end{equation}
Initial data for a collision are constructed by superposing two such configurations at $t=0$, centered at different positions and with opposite velocities, which can then be numerically evolved. For the non-rotating head-on mergers considered in this work, the resulting $1+1$ system can be solved efficiently using \textsl{Mathematica}'s \texttt{NDSolve}.

The only numerical difficulty arises from the singular behavior of the nonlinear term as $m\to 0$. In \cite{Andrade:2018yqu,Andrade:2019edf} this was addressed in two ways: imposing a small nonzero boundary condition $m(t,|z|=L)=m_{\rm IR}$ at the spatial boundary $|z|=L\gg 1$, which introduces a small background mass density; or modifying the equations by replacing $m\to |m-m_{\rm IR}|$ in the denominator of the nonlinear term. Both methods perform comparably, with the first behaving slightly more uniformly as $m_{\rm IR}\to 0$; we therefore adopt it as our regulator.

\begin{figure}[th]
    \centering
    \includegraphics[width=0.9\columnwidth]{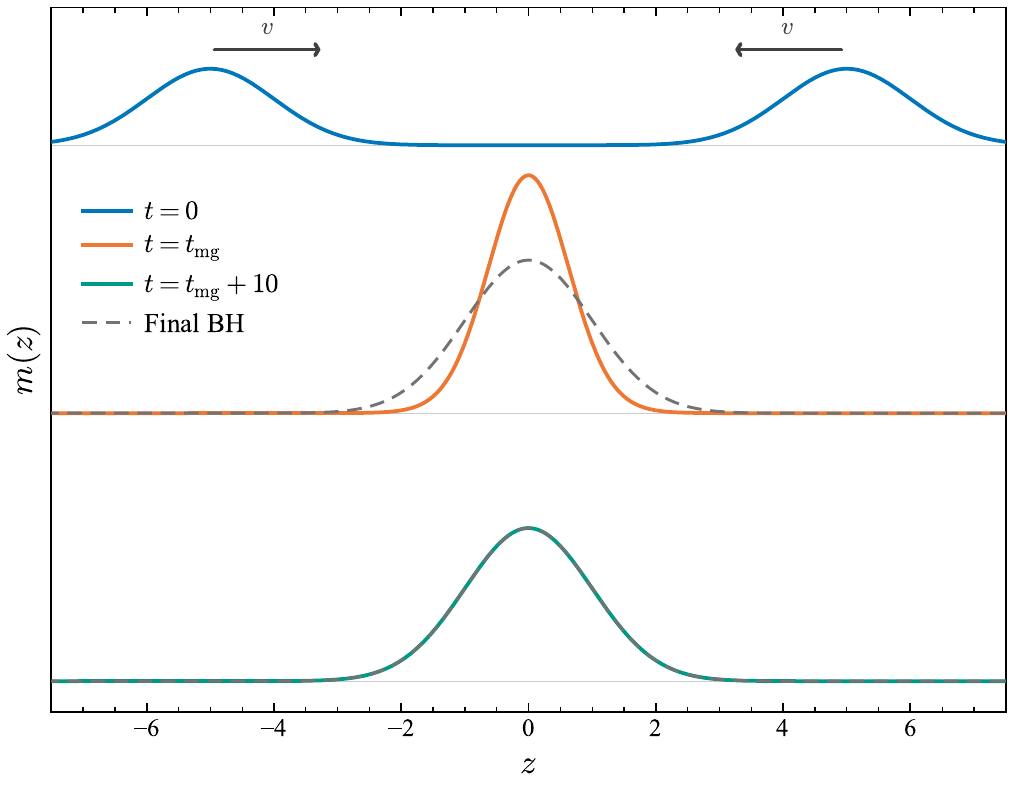}
    \caption{Head-on collision and merger of two black holes in the large-$D$ effective theory. The black holes are represented by Gaussian blobs of the mass density profile $m(z)$, initially moving toward each other with velocity $v$. We show the evolution for an equal-mass merger at three representative times: the initial configuration $t=0$, the merger time $t_{\rm mg}$, and a late stage of the evolution $t_{\rm mg}+10$. The dashed curves in the middle and bottom panels indicate the mass profile of the remnant black hole, determined by mass conservation, and illustrate the relaxation of the merged configuration toward equilibrium.}
    \label{fig:snapshots}
\end{figure}

We conclude with another important feature of the effective theory. Equations \eqref{eq:effective_eqs} imply that, for solutions in which $m$ and $p$ vanish as $z\to\pm\infty$ (or are periodic in $z$), the quantities $\int dz\, m(t,z)$ and $\int dz\, p(t,z)$ remain constant. In other words, the total energy and momentum of the system are conserved throughout the evolution.\footnote{The evolution is nevertheless irreversible and characterized by a non-decreasing entropy \cite{Andrade:2020ilm}.} This reflects the complete decoupling of gravitational radiation in the $1/D$ expansion. As a result, the ringdown signal is not extracted from gravitational waves measured at asymptotic infinity, but rather from the decay of the horizon profile encoded in the mass-density function $m(t,z)$. In the case of head-on collisions, these conservation laws entirely fix the properties of the final remnant black hole in terms of the initial data.

%%%%%%%%%%%%%%%%%%%%%%%%%%%%%%%%%%%%%%%%
\section{Black Hole Perturbation Theory}\label{sec:BHPT}
%%%%%%%%%%%%%%%%%%%%%%%%%%%%%%%%%%%%%%%%

The simplicity of the effective equations \eqref{eq:effective_eqs} and their black hole solution \eqref{eq:gblob} makes black hole perturbation theory in the large-$D$ limit analytically solvable to arbitrary order. Building on the linear analysis of~\cite{Andrade:2018nsz}, we develop an algorithm that systematically extends the construction to any perturbative order.

\subsection{Linear analysis and QNMs}

We start from the Gaussian blob solution for a single black hole, $(m_{\text{b}}(z), p_{\text{b}}(z))$ in \eqref{eq:gblob}, with unit amplitude, and perturb it as\footnote{$\epsilon$ is a smallness, bookkeeping parameter.}
\begin{equation}
    m=m_{\text{b}}(z) +\epsilon\, e^{-i\omega t}\delta m(z)\,,\quad p=p_{\text{b}}(z) +\epsilon\, e^{-i\omega t}\delta p(z)\,.
\end{equation}
Substituting into \eqref{eq:effective_eqs} and linearizing in the perturbations yields
\begin{equation}
\begin{aligned}\label{eq:pm}
    \delta m''-\delta p' +i\omega\, \delta m &=0\,,\\
    \delta p'' +2 z\, \delta p'+(z^2+1) \delta m'+(2+i\omega)\delta p +2 z\, \delta m&=0\,.
\end{aligned}
\end{equation}
We can eliminate $\delta p$ to obtain a fourth-order equation for $\delta m$. It is convenient to work with
\begin{equation}
    \delta\mathcal{R}(z)=\mb^{-1}(z) \delta m (z)\,,
\end{equation}
for which the equation takes the form
\begin{equation}\label{eq:4thR}
\begin{aligned}
     \delta\mathcal{R}^{(4)} -2z\,\delta\mathcal{R}^{(3)} +(2i\omega +z^2-1)\,\delta\mathcal{R}''& \\ -2i\omega z\,\delta\mathcal{R}' -\omega(\omega-2i)\,\delta\mathcal{R}&=0\,.
\end{aligned}
\end{equation}
Following \cite{Andrade:2018nsz} we factorize the differential operator as
\begin{equation}\label{eq:factor}
    \mathcal{L}_{h_{-}} \mathcal{L}_{h_{+}} \,\delta\mathcal{R}=0 \,,
\end{equation}
where
\begin{equation}\label{eq:L}
    \mathcal{L}_{h}=\partial_z^2-z \partial_z +h\,,
\end{equation}
and the constants $h_{\pm}$ are
\begin{equation}\label{eq:h_freqs}
    h_{\pm}(\omega)=\frac{1}{2}\left(1+2i\omega\pm\sqrt{1-4i\omega}\right)\, .
\end{equation}
Solutions are thus constructed by solving 
\begin{equation}\label{eq:delR_eq}
    \mathcal{L}_{h_{+}}\delta \mathcal{R}=0\,,
\end{equation}
which is the Hermite equation in the variable $z/\sqrt{2}$. Requiring bounded, finite polynomial solutions forces %\dpr{I sent $h\to h_{+}$ in the eq below. The reason is that $\omega_\ell = \pm \sqrt{\ell-1}-i(\ell-1)$ satisfies $h_{+}=\ell$ but \textit{not} $h_{-}=\ell$}
\begin{equation}\label{eq:quanth}
    h_{+}=\ell\in\mathbb{Z}_+\,. 
\end{equation}
The frequencies are then quantized as
\begin{equation}\label{eq:qnmfreq}
    \omega_\ell = \pm \sqrt{\ell-1}-i(\ell-1)\,,
\end{equation}
with corresponding wavefunctions
\begin{equation}\label{eq:qnm_wavefun}
    \delta \mathcal{R}=H_\ell\left( z/\sqrt{2}\right)\,.%\quad \delta m=e^{-z^2/2} H_n\left( z/\sqrt{2}\right)\,.
\end{equation}
The solution for $\delta p$ follows directly from \eqref{eq:pm}.

These solutions correspond to genuine quasinormal modes for $\ell\geq 2$, whereas $\ell=1$ is a translational zero mode. The integer $\ell$ can be identified with the partial-wave number of Schwarzschild perturbations, describing deformations of the sphere along a polar direction \cite{Emparan:2014aba,Bhattacharyya:2015dva}. The present effective theory captures only scalar (polar) perturbations; vector (axial) modes lie outside its $1+1$-dimensional description. As noted earlier, the isospectrality between polar and axial perturbations is lost for any $D>4$.

The simplicity of the problem allows us to go further and construct the Green's function governing arbitrary perturbations, not just individual quasinormal modes. For generic frequencies, the two independent solutions are
\begin{equation}
    \delta\mathcal{R}_{\pm}(z)=e^{z^{2}/4}D_{h_{+}}(\pm z)\, ,
\end{equation}
where $D_{\nu}(z)$ denotes a parabolic cylinder function. Their asymptotic behaviour is
\begin{equation}
    \delta\mathcal{R}_{\pm}(z)\overset{z\to\pm \infty}{\sim}z^{h_{+}}\, ,\qquad \delta\mathcal{R}_{\pm}(z)\overset{z\to\mp \infty}{\sim}\frac{e^{z^{2}/2}}{z^{1+h_{+}}}\,.
\end{equation}
Consequently, $\delta m_{\pm}$ is exponentially suppressed at one end of the geometry but not at the other. %\dpr{actually $\delta m_{\pm}$ does not diverge exponentially at the other end, it is simply not exponentially suppressed, by looking at the second of the eqs above and recalling $\delta m_{\pm}\sim e^{-z^{2}/2}\delta\mathcal{R}_{\pm}(z)$}.
Quasinormal modes are singled out by requiring exponential decay as $|z|\to\infty$ on both sides. This occurs precisely when $h_+$ is a positive integer, in which case the parabolic cylinder functions reduce to finite polynomials, as we saw above.

The corresponding Green's function for $e^{-z^{2}/4}\delta\mathcal{R}(z)$ is
\begin{equation}\label{eq:prop}
    G(z,z')=\frac{\Gamma(-h_{+})}{\sqrt{2\pi}}D_{h_{+}}(z_{>})D_{h_{+}}(-z_{<})\, ,
\end{equation}
where $z_{>}=\text{max}(z,z')$, $z_{<}=\text{min}(z,z')$ and we have used the Wronskian relation
\begin{equation}
    W\left[D_{\nu}(z),D_{\nu}(- z)\right]=\frac{\sqrt{2\pi}}{\Gamma(-\nu)}\,.
\end{equation}
The quasinormal modes correspond to poles of the Green's function, or equivalently to zeros of the Wronskian, which again occur when $h_+$ is a positive integer (Eq.~\eqref{eq:quanth}), thus confirming the previous analysis.

Furthermore, the quasinormal modes exhaust the entire response of the propagator. Unlike the four-dimensional case, the frequency dependence of \eqref{eq:prop} contains no branch cuts and therefore gives rise to no power-law tails. In four dimensions, such tails arise from branch cuts associated with the backscattering of gravitational waves at large distances \cite{Leaver_prop}. Their absence here is expected, since this far-zone physics is entirely removed in the large-$D$ near-horizon limit.

Collecting the results above, the real-valued QNM solutions for generic $\ell\geq 2$ are given by
\begin{widetext}
\begin{equation}\label{eq:realQNMs}
    \begin{aligned}
        \delta m(t,z)&=A \times H_{\ell }\left(\frac{z}{\sqrt{2}}\right) e^{-t \ell +t-\frac{z^2}{2}} \cos \left(t \sqrt{\ell -1}+\phi \right)\, ,\\
        \delta p(t,z)&=-A\times\frac{e^{-t \ell +t-\frac{z^2}{2}}}{\ell +3}\Biggl(4 \sqrt{2} \sqrt{\ell -1} H_{\ell -1}\left(\frac{z}{\sqrt{2}}\right) \sin \left(t \sqrt{\ell -1}+\phi \right)\\
        & +\left(z (\ell +3) H_{\ell }\left(\frac{z}{\sqrt{2}}\right)+\sqrt{2} ((\ell -1) \ell -4) H_{\ell -1}\left(\frac{z}{\sqrt{2}}\right)\right) \cos \left(t \sqrt{\ell -1}+\phi \right)\\
        &-2 z (\ell -1) H_{\ell -2}\left(\frac{z}{\sqrt{2}}\right) \left((\ell +1) \cos \left(t \sqrt{\ell -1}+\phi \right)-\sqrt{\ell -1} \sin \left(t \sqrt{\ell -1}+\phi \right)\right)\\
        &+2 \sqrt{2} (\ell -2) (\ell -1) H_{\ell -3}\left(\frac{z}{\sqrt{2}}\right) \left((\ell +1) \cos \left(t \sqrt{\ell -1}+\phi \right)-\sqrt{\ell -1} \sin \left(t \sqrt{\ell -1}+\phi \right)\right)\Biggr)\,,
    \end{aligned}
\end{equation}
\end{widetext}
with arbitrary amplitude $A$ and phase $\phi$.

This completes the first-order, linear analysis. The simplicity of the effective theory allows the construction to be extended systematically to nonlinear orders. We now show how a QNM solution known up to order $N-1$ determines the solution at order $N$ recursively.

\subsection{Nonlinear QNMs at $N$-th Order}

Consider a perturbative solution of the form
\begin{equation}\label{eq:Nth_QNM}
\begin{aligned}
    m(t,z)&=m_{b}(t,z)+\sum_{i=1} \epsilon^i  \delta^i m(t,z)\,,\\
    p(t,z)&=p_{b}(t,z)+\sum_{i=1} \epsilon^i  \delta^i p(t,z)\, ,
\end{aligned}
\end{equation}
where $\epsilon$ is again a bookkeeping parameter. By an \textit{$N$th-order QNM} we mean the solution truncated at order $\epsilon^N$, assuming that the first-order perturbation is a quasinormal mode (or a superposition thereof) of the form \eqref{eq:realQNMs}.
% where $\epsilon$ is a book-keeping parameter measuring the smallness of the perturbation. We refer as the \textit{$N$th-order QNM} to the solution \eqref{eq:Nth_QNM} up to order $N$, where the linear ($N=1$) solution is a (combination of) QNMs given in \eqref{eq:realQNMs}.

The equations governing the $N$th-order perturbations are
\begin{equation}\label{eq:Nthorder_eq_TD}
\begin{aligned}
    &\delta^{N} m_t+\delta^{N} p_z - \delta^{N} m_{zz} =0\,,\\  \\
    &\delta^{N} p_t-2z\,\delta^{N} m - 2\,\delta^{N} p- (1 + z^2)\delta^{N} m_z\\&
    - 2z\,\delta^{N} p_z- \delta^{N} p_{zz}=S_{N}\, ,
\end{aligned}
\end{equation}
where the source $S_N$ depends only on lower-order perturbations, $\delta^M m$ and $\delta^M p$ with $M<N$. Assuming that the linear solution is a superposition of the QNMs \eqref{eq:realQNMs}, the nonlinear source can be decomposed as
\begin{equation}\label{eq:s_expansion}
\begin{aligned}
    S_{N}(t,z)&=\sum_{k}\left(e^{-i\sigma_{N,k}t}S_{N,k}(z)+c.c.\right)\\&+\sum_{k_{\circ}}e^{-i\sigma_{N,k_{\circ}}t}S_{N,k_{\circ}}(z)\,,
\end{aligned}
\end{equation}
where the frequencies $\sigma_{N,k}$ satisfy $\mathrm{Re}(\sigma_{N,k})\neq0$, while $\mathrm{Re}(\sigma_{N,k_\circ})=0$. These correspond, respectively, to oscillatory and purely damped nonlinear channels at order $N$.

As a simple example, if the linear perturbation consists of a single QNM with frequency $\omega$, then at quadratic order one finds an oscillatory channel with frequency $\sigma_{2,1}=2\omega$ and a purely damped channel with frequency $\sigma_{2,\circ}=\omega-\bar{\omega}$. Although the decomposition \eqref{eq:s_expansion} is expressed in terms of complex exponentials for convenience, the complete source is of course real. Since $S_N$ depends nonlinearly on lower-order perturbations, it is essential to construct it from the genuine real solutions; otherwise one generates spurious frequency channels.
% for some frequencies $\sigma_{N,k}$ with $\text{Re}(\sigma_{N,k})\ne0$, and others $\sigma_{N,k_{\circ}}$ with $\text{Re}(\sigma_{N,k_{\circ}})=0$. These define, respectively, the nonlinear oscillatory and purely damped channels at order $N$. For example, considering that the linear solution is a single QNM with frequency $\omega$, then at quadratic order one has an oscillatory channel $\sigma_{2,1}=2\omega$ and a purely damped one, $\sigma_{2,\circ}=\omega-\bar{\omega}$. Such an expansion in terms of complex exponentials is convenient for computations, but it is clear the sum yields a real result. We also note that since $S_{N}(t,z)$ depends nonlinearly on the lower order terms of \eqref{eq:Nth_QNM}, it is required that one uses the genuine real lower order solutions to compute it -- otherwise one would pick spurious terms.

Once the nonlinear channels have been identified, one may seek solutions of the form
\begin{equation}
    \begin{aligned}
    \delta^{N} m(t,z)&=\sum_{k}\left(e^{-i\sigma_{N,k}t}\ \delta^{N} m_{k}(z)+c.c.\right)\\
    &\quad +\sum_{k_{\circ}}e^{-i\sigma_{N,k_{\circ}}t}\ \delta^{N} m_{k_{\circ}}(z)\, ,\\ %\notag \\
    \delta^{N} p(t,z)&=\sum_{k}\left(e^{-i\sigma_{N,k}t}\ \delta^{N} p_{k}(z)+c.c.\right)\\
    &\quad +\sum_{k_{\circ}}e^{-i\sigma_{N,k_{\circ}}t}\ \delta^{N} p_{k_{\circ}}(z)\, ,
\end{aligned}
\end{equation}
and solve the equations independently in each channel.

This procedure constructs a particular solution of \eqref{eq:Nthorder_eq_TD}, namely the one sourced entirely by lower-order perturbations and ultimately by the linear QNM. It is this contribution that we identify as the $N$th-order QNM. The homogeneous solutions at order $N$ are identical to the linear modes and correspond to independent excitations---for example those associated with initial data---which are not part of the recursive construction considered here.
%and proceed channel by channel. Of course, this only provides a particular solution to \eqref{eq:Nthorder_eq_TD}, the one that is purely driven by the source term built out of lower order solutions (and ultimately by the linear QNM). This is the part of the solution that is regarded as the $N$th order QNM. Solutions to the homogeneous problem at $N$th order are identical to the linear solutions, and are driven by other effects (e.g. initial data) that we are not considering in this construction. 

The solution in each channel can be constructed as follows. Repeating the linear analysis but now retaining the source term $S_N$, one obtains a decoupled equation for $\delta^{N}\mathcal{R}_{k}(z)$, defined by
\begin{equation}
    \delta^{N}m_{k}(z)=m_{b}(z)\delta^{N}\mathcal{R}_{k}(z)\, ,
\end{equation}
which takes the form
\begin{equation}\label{eq:RN}
    \mathcal{L}_{h_{-}}\mathcal{L}_{h_{+}}\delta^{N}\mathcal{R}_{k}(z)=-e^{z^{2}/2}\partial_{z}S_{N,k}(z)\equiv\mathcal{S}_{N,k}(z)\, ,
\end{equation}
where $\mathcal{L}_h$ is defined in \eqref{eq:L}, and $h_{\pm}$ are given by \eqref{eq:h_freqs} evaluated at the nonlinear frequencies $\sigma_{N,k}$,
\begin{equation}
    h_{\pm}(\sigma_{N,k})=\frac{1}{2}\left(1+2i\sigma_{N,k}\pm\sqrt{1-4i\sigma_{N,k}}\right)\, .
\end{equation}
For the present construction, the source $\mathcal{S}_{N,k}(z)$ is a polynomial whose degree scales as $\deg \mathcal{S} = N \deg \delta\mathcal{R}$, and can therefore be expanded in a \emph{finite} basis of Hermite polynomials,
\begin{equation}
    \mathcal{S}_{N,k}(z)=\sum_{\ell}s^{\ell}_{N,k}H_{\ell}(z/\sqrt{2})\, ,
\end{equation}
with constant coefficients $s^{\ell}_{N,k}$. Moreover, the Hermite polynomials diagonalize the operator $\mathcal{L}_{h}$, satisfying
\begin{equation}
    \mathcal{L}_{h_{-}}\mathcal{L}_{h_{+}}H_{\ell}(z/\sqrt{2})=(h_{-}-\ell)(h_{+}-\ell)H_{\ell}(z/\sqrt{2})\, .
\end{equation}
It follows immediately that
\begin{equation}
    \delta^{N}\mathcal{R}_{k}=\sum_{\ell}\frac{s^{\ell}_{N,k}}{(h_{-}-\ell)(h_{+}-\ell)}H_{\ell}(z/\sqrt{2})\, .
\end{equation}
From this expression, $\delta^{N}m_{k}(z)$ follows directly, while $\delta^{N}p_{k}(z)$ is obtained from the second equation in \eqref{eq:Nthorder_eq_TD} using the first equation, which yields
\begin{equation}\label{eq:p_recon}
\begin{aligned}
&(\sigma_{N,k} - 2i)\,\delta^{N}p_{k}(z)
= 2iz(1 + i\sigma_{N,k})\delta^{N}m_{k}(z) \\
& + i(1 + z^2 + i\sigma_{N,k})\delta^{N}m_{k,z}(z) + 2iz\,\delta^{N}m_{k,zz}(z)\\
&+ i\,\delta^{N}m_{k,zzz}(z) - \frac{\sigma_{N,k} - 2i}{2+i\sigma_{N,k}}\, S_{N,k}(z)\, .
\end{aligned}
\end{equation}
This completes the construction of the solution at $N$th order, given all lower-order data. It is convenient to introduce the amplitudes
\begin{equation}
\begin{aligned}
     A^{\ell}_{N,k}&\equiv\frac{s^{\ell}_{N,k}}{(h_{-}(\sigma_{N,k})-\ell)(h_{+}(\sigma_{N,k})-\ell)}\,,\\  A^{\ell}_{N,k_{\circ}}&\equiv\frac{s^{\ell}_{N,k_{\circ}}}{(h_{-}(\sigma_{N,k_\circ})-\ell)(h_{+}(\sigma_{N,k_\circ})-\ell)}\, ,
\end{aligned}
\end{equation}
with $A^{\ell}_{N,k}$ complex and $A^{\ell}_{N,k_{\circ}}$ real. The resulting expression for $\delta^{N} m(t,z)$ can then be written as
\begin{equation}
\begin{aligned}\label{eq:dm_channels}
    \frac{\delta^{N}m(t,z)}{m_{b}(z)}&= \sum_{\ell}\Biggl\{\Bigl[\sum_{k}\left(A^{\ell}_{N,k}e^{-i\sigma_{N,k}t}+c.c.\right)\\&+\sum_{k_\circ}A^{\ell}_{N,k_{\circ}}e^{-i\sigma_{N,k_\circ}t}\Bigr]H_{\ell}\left(z/\sqrt{2}\right)\Biggr\}\,.
\end{aligned}
\end{equation}
These amplitudes quantify the excitation of each nonlinear channel within a given ``angular level'' $\ell$, and are fully determined by the parameters of the linear QNMs. In particular, they are analytic functions of the linear QNM amplitudes and phases. In this sense, the excitation strengths of nonlinear QNMs are predictive outputs of the theory, on equal footing with the QNM frequencies themselves.

The algorithm presented here is remarkably simple. This simplicity stems from the fact that the linearized problem is governed by a second-order operator admitting a complete set of polynomial eigenfunctions. A similar structure played a key role in the derivation of nonlinear eikonal QNMs from the Penrose limit in~\cite{Fransen:2025cgv}. By contrast, the Teukolsky operator governing black hole perturbations in $D=4$ does not admit an exact polynomial eigenbasis analogous to the Hermite functions used here. Nevertheless, it can be brought into a tridiagonal form in a suitable polynomial basis~\cite{London:2023idh}, which partially recovers this algebraic structure.

Next we present a set of explicit solutions that will be useful for comparison with the numerical simulations in Section~\ref{sec:Head_On_Black_Hole_Mergers}.

%%%%%%%%%%%%%%%%%%%%%%%%%%%%%%%%%%%%%%%%%%%%%
\subsection{Explicit solutions}\label{sec:Explicit_solutions}
%%%%%%%%%%%%%%%%%%%%%%%%%%%%%%%%%%%%%%%%%%%%%

From \eqref{eq:dm_channels}, the full order-$N$ solution admits an expansion in a finite Hermite basis,
\begin{equation}\label{eq:Hermites}
\begin{aligned}
    \delta^{N}m(t,z)&\equiv m_{b}(z)\sum_{\ell}\mathcal{R}_{N}^{\ell}(t)H_{\ell}(z/\sqrt{2})\, ,\\
    \delta^{N}p(t,z)&\equiv m_{b}(z)\sum_{\ell}p^{\ell}_{N}(t)H_{\ell}(z/\sqrt{2})\, .
\end{aligned}
\end{equation}
The solution is therefore fully characterized by the time-dependent coefficients $\mathcal{R}_{N}^{\ell}(t),p^{\ell}_{N}(t)$, which are obtained through the construction described above. In what follows, we focus on $\mathcal{R}_{N}^{\ell}(t)$, since these coefficients are directly related to the observables extracted in the numerical simulations.

\subsubsection{Second-order Solutions}

Assuming a single linear QNM with $\ell=2$ (and therefore frequency $\omega=\pm 1-i$), of amplitude and phase $A$ and $\phi$, one finds that at second order the response splits into an oscillatory and a purely damped channel, with
\begin{equation}\label{eq:sigma_Selfcoupling_secondorder}
    \sigma_{2,1}=2-2i\, ,\quad \sigma_{2,\circ}=-2i\, .
\end{equation}
The corresponding amplitudes are
\begin{equation}\label{eq:A_Selfcoupling_secondorder}
\begin{aligned}
      &A^{2}_{2,1}=A^{2}\left(-\frac{1}{5}+\frac{i}{10}\right) e^{-2 i \phi }\, ,\quad A^{4}_{2,1}=A^{2}\frac{1}{8} e^{-2 i \phi }\,, \\
        &A^{2}_{2,\circ}=A^{2}\, ,\quad A^{4}_{2,\circ}=A^{2}\frac{1}{4}\, ,
\end{aligned}
\end{equation}
and the mass coefficients take the form
 \begin{equation}
     \begin{aligned}\label{eq:m_Selfcoupling_secondorder}
     \mathcal{R}^{2}_{2}(t)&=-A^{2}\frac{1}{5} e^{-2 t} (-\sin (2 t+2 \phi )+2 \cos (2 t+2 \phi )-5)\, ,\\
     \mathcal{R}^{4}_{2}(t)&=A^{2}\frac{1}{4} e^{-2 t} (\cos (2 t+2 \phi )+1)\, .
\end{aligned}
 \end{equation}
This construction generalizes straightforwardly to nonlinear QNMs sourced by two linear modes, for instance with $\ell=2$ and $\ell=3$, corresponding to frequencies $\omega_{2}=\pm1-i$ and $\omega_{3}=\pm\sqrt{2}-2i$, and arbitrary amplitudes $A_{2},A_{3}$ and phases $\phi_{2},\phi_{3}$. The explicit expressions are given in Appendix~\ref{ap:n2n3}.

\subsubsection{Third-order Solutions}

Extending the self-coupling of a single $\ell=2$ QNM to third order and using the second-order solution \eqref{eq:m_Selfcoupling_secondorder}, we find two oscillatory channels and no purely damped contribution,
    \begin{equation}
        \sigma_{3,1}= 1-3i\,, \quad \sigma_{3,2}= 3-3i\,.
    \end{equation}
The corresponding amplitudes are
    \begin{equation}
        \begin{aligned}
        &A^{2}_{3,1}=A^{3}\left(-\frac{3}{20}+\frac{19 i}{20}\right) e^{-i \phi }\,,\quad A^{4}_{3,1}=A^{3}\left(\frac{2}{5}+\frac{i}{20}\right) e^{-i \phi }\, , \\ & A^{6}_{3,1}=A^{3}\frac{e^{-i \phi }}{16}\, ,\quad A^{2}_{3,2}=A^{3}\left(\frac{3}{20}-\frac{3 i}{20}\right) e^{-3 i \phi }\,,\\
        & A^{4}_{3,2}=A^{3}\left(-\frac{1}{10}+\frac{i}{20}\right) e^{-3 i \phi }\,,\quad A^{6}_{3,2}=A^{3}\frac{1}{48} e^{-3 i \phi }\,,
    \end{aligned}
    \end{equation}
and the Hermite coefficients $\mathcal{R}^{\ell}_{3}(t)$ are
     \begin{widetext}
         \begin{equation}\label{eq:third_order_ms}
         \begin{aligned}
             \mathcal{R}^{2}_{3}(t)&=-A^{3}\frac{1}{10} e^{-3 t} (-19 \sin (t+\phi )+3 \sin (3 t+3 \phi )+3 \cos (t+\phi )-3 \cos (3 t+3 \phi ))\, ,\\
             \mathcal{R}^{4}_{3}(t)&=A^{3}\frac{1}{10} e^{-3 t} (\sin (t+\phi )+\sin (3 t+3 \phi )+8 \cos (t+\phi )-2 \cos (3 t+3 \phi ))\,,\\
             \mathcal{R}^{6}_{3}(t)&=A^{3}\frac{1}{24} e^{-3 t} (3 \cos (t+\phi )+\cos (3 t+3 \phi ))\, .
         \end{aligned}
     \end{equation}
     \end{widetext}
Notice that the pseudo-resonant channel $\sigma_{3,1}$ is parametrically enhanced relative to the other contributions, with $|A^2_{3,1}|/A^3 \sim \mathscr{O}(1)$. This suggests that cubic QNM effects, although suppressed by the amplitude of the initial perturbation, may nevertheless have observable consequences in black hole ringdown~\cite{Dyer:2025hdt}.
%

%%%%%%%%%%%%%%%%%%%%%%%%%%%%%%%%%%%%%%%%%%%%%
\section{Head-On Black Hole Mergers}\label{sec:Head_On_Black_Hole_Mergers}
%%%%%%%%%%%%%%%%%%%%%%%%%%%%%%%%%%%%%%%%%%%%%

Black hole mergers in the large-$D$ limit retain the familiar qualitative picture: after coalescence, the system relaxes to equilibrium through a characteristic ringdown governed by the linear QNM spectrum. At the same time, the effective description differs in important ways from finite-$D$ gravity. In particular, in the effective large-$D$ theory \eqref{eq:effective_eqs}, black holes are not described as sharply localized objects, but as Gaussian blobs that remain continuously connected to the brane horizon, so a global notion of merger time is absent. Moreover, the dynamics captured by the effective equations describes only the near-horizon region, which decouples from asymptotic radiation in the large-$D$ limit.

We focus on the simplest setup: head-on collisions of non-rotating, nearly equal-mass black holes. While more general configurations—including spinning and non-head-on mergers—have been studied previously~\cite{Andrade:2019edf}, this restricted setting provides a clean arena in which to isolate nonlinear effects in the post-merger relaxation. In what follows, we use it as a controlled laboratory to probe the emergence of nonlinearities in ringdown, with extensions to very unequal masses and rotation left to future work.

\subsection{Numerics}

\begin{figure}
    \centering
    \includegraphics[width=\columnwidth]{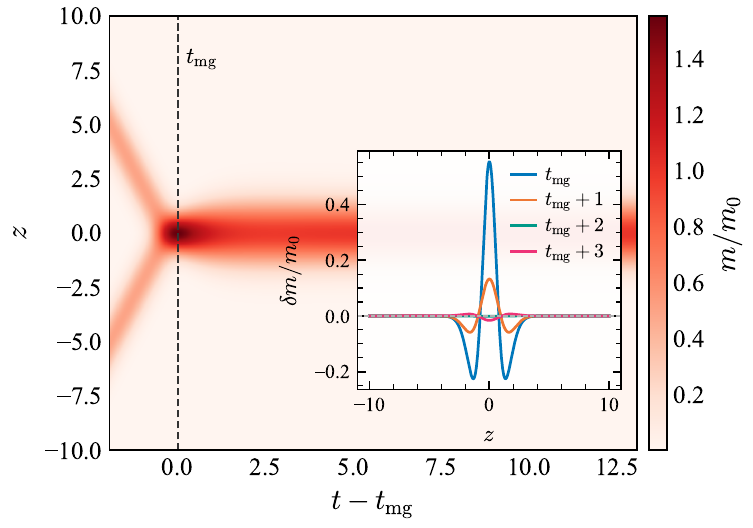}
    \caption{ Evolution of $m(t,z)$ during an equal mass merger $q=1$ with $v=3$. The dashed, black line indicates the reference time $t_{\rm mg}$. The inset shows the relaxation of the final black hole to equilibrium, $\delta m = m(t,z) - m_{\rm rem}(z)$, at different times, as indicated in the legend.}
    \label{fig:merger}
\end{figure}

We have seen that the effective theory admits solutions describing a single isolated black hole moving at constant velocity, as in \eqref{eq:movingblob}. More generally, the solutions can be translated and boosted to give
\begin{equation}
    \begin{aligned}
        m_{v,z_0} =& \mb(z-z_0-vt) = m_0 e^{-(z-z_0-vt)^2/2} \, , \\
        p_{v,z_0} =& \Bigl[v_0 - (z-z_0-vt)\Bigr]\mb(z) \,.
    \end{aligned}
\end{equation}
We construct initial data describing two black holes moving towards each other by superposing two such configurations,
\begin{equation}
    \begin{aligned}
           m(0,z) =& m_{v,-qz_0} + q\, m_{-v/q, z_0} \, , \\ 
           p(0,z) =& p_{v,-qz_0} + q\, p_{-v/q,z_0} \, , 
    \end{aligned}
\end{equation}
where $q$ denotes the mass ratio, with $q=1$ corresponding to equal-mass mergers.

We evolve the equations numerically on a finite domain $z\in[-L/2,L/2]$, with boundary conditions
\begin{equation}
    m(t,\pm L/2) = m_{\rm IR} \, , \qquad p(t, \pm L/2) = 0 \, .
\end{equation}
As discussed in Section~\ref{sec:BHDyn}, we introduce a small regulator, typically $m_{\rm IR}=10^{-6}$, to avoid numerical issues associated with the nonlinear denominator in eq.~\eqref{eq:effective_eqs}. This regulator limits the accuracy of the solution at amplitudes of order $m_{\rm IR}$ and, in particular, can affect the extraction of rapidly damped higher harmonics. On the other hand, it significantly simplifies and accelerates the numerical implementation. Alternative numerical implementations of the effective equations are possible, see e.g.~\cite{Andrade:2019edf}.

Since the total mass is conserved during the evolution, the final remnant takes the form
\begin{equation}\label{eq:remnant}
    m_{\rm rem} = m_0(1+q) e^{-(z-v_{\rm kick}t)^2/2} \,  ,
\end{equation}
where the kick velocity $v_{\rm kick}\ll 1$ is small and can be neglected for mergers with $q \gtrsim 1/2$. An example of the merger and subsequent relaxation to equilibrium is shown in Fig.~\ref{fig:merger}. Although a precise notion of merger time is not available—since black holes are not sharply localized—we define $t_{\rm mg}$ as the time at which $m(t,0)$ attains its maximum. This is indicated by the dashed line in the left panel of Fig.~\ref{fig:merger}. The inset panel shows that, after coalescence, the horizon deformation rapidly relaxes to the stationary solution $m_{\rm rem}$.

\subsection{Extracting Ringdown Modes}

\begin{figure}
    \centering
    \includegraphics[width=\columnwidth]{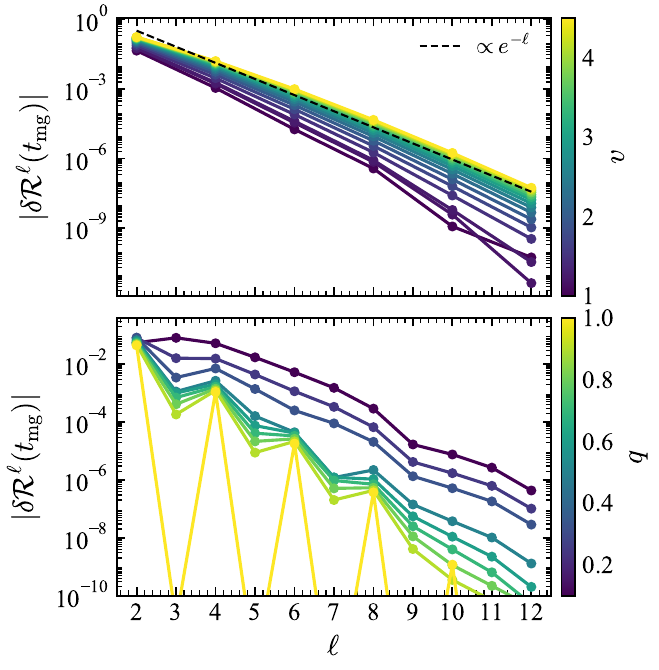}
    \caption{Amplitude of $\delta\mathcal{R}^\ell$ for the different multipole numbers $\ell$ as a function of the merger velocity (\textbf{top}), for equal mass mergers, $q=1$; and as a function of the mass ratio (\textbf{bottom}), setting $v=1$. Even at high velocities, the higher harmonics are exponentially suppressed. On the other hand, in the limit $q \to 0$ we observe a flattening of the spectrum, in agreement with perturbative calculations up to $D=10$~\cite{Berti:2003si} (see also \cite{Berti:2010gx}). }
    \label{fig:spectrum}
\end{figure}

From the perturbative analysis, we expect the horizon relaxation to decompose in Hermite modes, governed by a hierarchy of damped oscillations. We define (cf.~\eqref{eq:Hermites})
\begin{equation}
    \delta\mathcal{R}^\ell = \frac{1}{N_\ell } \int_{-\Lambda}^{\Lambda}\left(m(t,z)-m_{\rm rem}(z)\right)H_\ell\left(\frac{z}{\sqrt2}\right) dz \, ,
\end{equation}
with normalization 
\begin{equation}
    N_\ell = 2^\ell \ell! \sqrt{2\pi}\, m_0(1+q)\,.
\end{equation}
The cutoff $\Lambda<L$ is introduced in the projection to mitigate the contamination from the numerical ``thermal atmosphere'' $m_{\rm IR}$. 

For equal-mass mergers, higher Hermite harmonics are exponentially suppressed, $\delta \mathcal{R}^\ell\sim e^{-\ell}$, with little dependence on the collision velocity (see Fig.~\ref{fig:spectrum}). Reflection symmetry about $z=0$ further implies that only even-$\ell$ modes are excited. As shown in the bottom panel of Fig.~\ref{fig:spectrum}, breaking this symmetry via mass asymmetry activates odd modes, which become comparable in magnitude to the even ones already for a mass ratio of $1:10$.

Due to the suppression of higher harmonics and the rapid decay of modes with $\ell>2$ (see eq.~\eqref{eq:qnmfreq}), we focus on modelling the $\ell=2$ multipole. Higher multipoles can be incorporated straightforwardly along the same lines. We consider three competing models, each including successively higher perturbative orders.
\begin{enumerate}
    \item \textbf{Linear QNMs (L).} The simplest model describes the ringdown in terms of a single linear QNM,
    \begin{equation}
        \delta\mathcal{R}_{(L)}^2 = A \cos(t+\phi)e^{-t} \, .
    \end{equation}
    It depends on two free parameters: an amplitude $A$ and a phase $\phi$, while the frequency and damping rate are fixed to their perturbative values \eqref{eq:qnmfreq}. We have also verified that allowing these quantities to vary reproduces the perturbative predictions with high accuracy.
    \item \textbf{Quadratic QNMs (Q).} This model includes second-order effects,
    \begin{equation}
        \begin{aligned}\label{eq:quadratic_model}
            \delta\mathcal{R}_{(Q)}^2 =& A \cos(t+\phi)e^{-t} -\frac{A^2e^{-2t}}{5}\\
            &\times \Bigl[2\cos(2t+2\phi)-\sin(2t+2\phi)-5\Bigr] \, .
        \end{aligned}
    \end{equation}
    This includes the linear QNM together with the quadratic response in \eqref{eq:m_Selfcoupling_secondorder}, with frequencies $\sigma_{2,1}$ and $\sigma_{2,\circ}$ from \eqref{eq:sigma_Selfcoupling_secondorder}. The model still has only two free parameters, since the quadratic amplitudes are fixed by the  second-order perturbative calculation.

    \item \textbf{Free Quadratic QNMs (Q*).} 
    In this model we relax this constraint and treat the quadratic amplitudes as free parameters,
    \begin{equation}
        \begin{aligned}
            \delta\mathcal{R}_{(Q*)}^2 =& A \cos(t+\phi)e^{-t} +A^2e^{-2t}\\
            &\times \Bigl[Q_+\cos(2t+2\phi)+Q_-\sin(2t+2\phi)+Q_\circ\Bigr] \, .
        \end{aligned}
    \end{equation}
    This introduces five free parameters in total. It provides a direct consistency check of the quadratic QNM prediction for the nonlinear amplitudes.
    %Thus, this model has $5$ free parameters. This will be useful to demonstrate that the amplitude of the quadratic QNMs is in excellent agreement with our perturbative calculation, serving as a consistency check. 
    %
    \item \textbf{Cubic QNMs (C).} Finally, we include third-order effects. Although subleading in amplitude, these terms contain contributions at the same frequency as the linear mode, making them potentially relevant for accurately modelling high-velocity, equal-mass mergers. Adding \eqref{eq:third_order_ms} to the lower-order terms yields
    %Finally, we can also include third-order effects. These are quite interesting since, even though they are short-lived, they contribute at the same (real) frequency as the linear QNMs. We anticipate that their inclusion will be important to model high-velocity, equal-mass mergers. The third-order contribution is given by \eqref{eq:third_order_ms}, which added to the linear and quadratic solutions gives the full cubic model model 
    %
    \begin{equation}
        \begin{aligned}
            \delta\mathcal{R}_{(C)}^2 =& A \cos(t+\phi)e^{-t} -\frac{A^2e^{-2t}}{5}\\
            &\times \Bigl[2\cos(2t+2\phi)-\sin(2t+2\phi)-5\Bigr] \\
            &-\frac{A^3 e^{-3t}}{10} \Bigl[-19 \sin (t+\phi )+3 \sin (3 t+3 \phi )\\
            &+3 \cos (t+\phi )-3 \cos (3 t+3 \phi )\Bigr] \, .
        \end{aligned}
    \end{equation}
    As before, only the amplitude and phase of the linear mode are free parameters; all nonlinear contributions are fixed by the perturbative structure of the theory.
\end{enumerate}

We determine the best-fit parameters for each model by minimizing the $L_2$ norm of the residual between the numerical signal and the template, using a standard least-squares routine implemented in \texttt{scipy}~\cite{virtanen2020scipy}.

To quantify the goodness of the fit, we compute the mismatch
\begin{equation}\label{eq:mismatch}
    \mathcal{M} = \frac{| \delta \mathcal{R}_2(t)-\delta\mathcal{R}_2^{(M)}(t)|^2}{|\delta\mathcal{R}_2(t)|^2} \, , 
\end{equation}
where $|\cdot|$ denotes the $L_2$ norm over the fitting interval and $\delta \mathcal{R}_2^{(M)}$ is the corresponding model evaluated at the best-fit parameters. The fit is performed over a finite time window $t\in[t_0,t_{\rm end}]$, with $t_{\rm end}\gtrsim 10$, i.e. typically more than ten e-folds of decay of the fundamental mode.

As a consistency check, we verify that the inferred amplitudes and phases are insensitive to the choice of $t_0$.

\subsection{Evidence of Nonlinearities}

\begin{figure}
    \centering
    \includegraphics[width=\columnwidth]{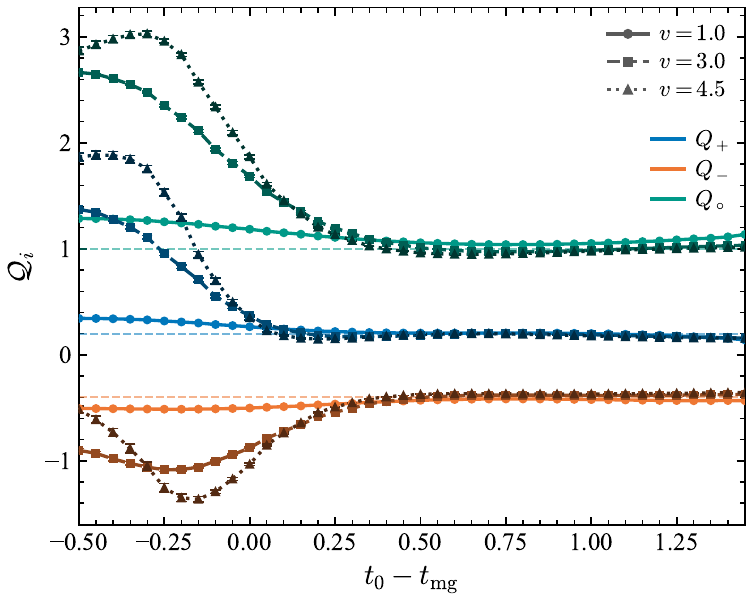}
    \caption{Extracted value of the quadratic QNM amplitude $Q_{\pm,\circ}$ as a function of the ringdown start time for three different simulations, as indicated in each panel. The dashed lines show the prediction from black hole perturbation theory~\eqref{eq:quadratic_model}. The inferred values of the quadratic QNM ratio converge to the perturbative calculations once we start fitting at sufficiently late times.}
    \label{fig:qqnms}
\end{figure}

In the following, we restrict to equal-mass mergers ($q=1$), for which contributions from $\ell\neq 2$ can be safely neglected. As a first step, we analyze the Q* model in order to test whether it reproduces the perturbative amplitudes of the quadratic QNMs in eq.~\eqref{eq:m_Selfcoupling_secondorder}. We consider three mergers with different collision velocities, and summarize the results in Fig.~\ref{fig:qqnms}. The fitted coefficients $Q_i$ ($i=\pm,\circ$) converge rapidly to the theoretical predictions in \eqref{eq:quadratic_model}. The agreement is particularly good at lower velocities, while deviations increase as the velocity grows. As discussed below, this is expected: at higher velocities, cubic effects become relevant and are not captured by the Q* model. Overall, this confirms that quadratic QNMs are excited in black hole ringdown in the large-$D$ limit, with amplitudes in excellent agreement with perturbative predictions.

\begin{figure*}
    \centering
    \includegraphics[width=\linewidth]{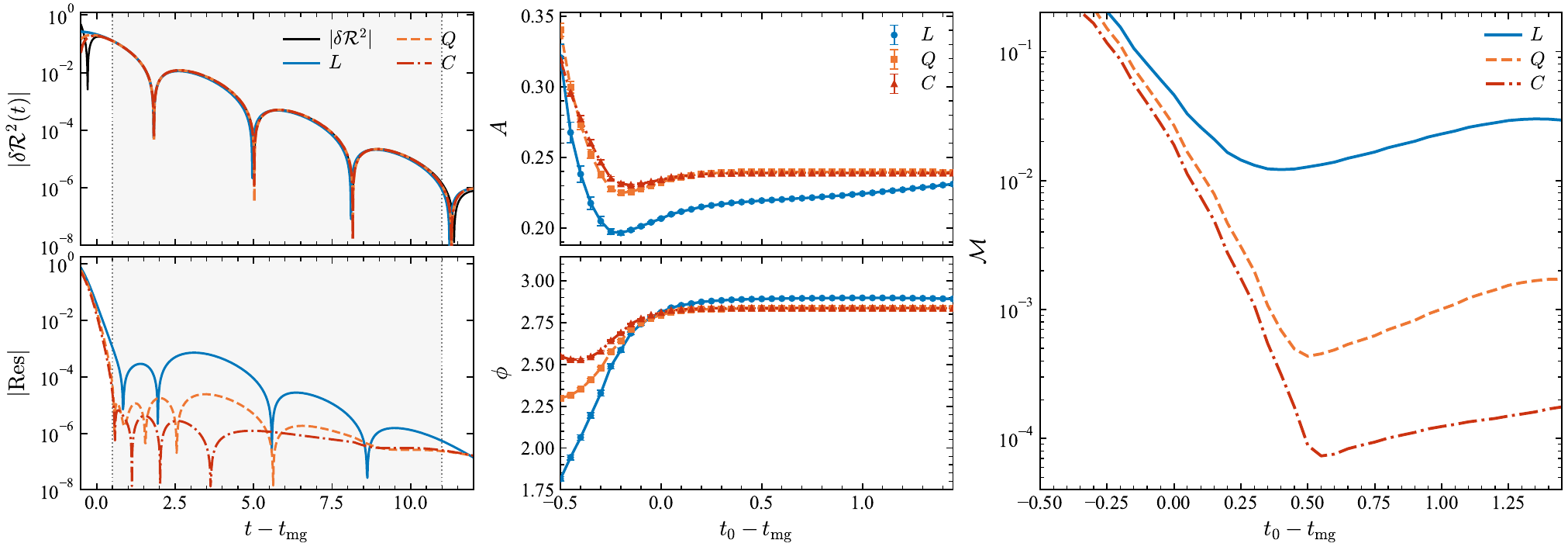}
    \caption{\textbf{Left:} Signal (black) and best fit using the L, Q, and C models (see legend), for an equal mass black hole merger with $v=4.5$, with a fit start time $t_0-t_{\rm mg}=0.5$. The bottom panel shows the residual, which clearly improves when including nonlinear effects. \textbf{Center:} Amplitude (top) and phase (bottom) of the linear QNM as a function of the ringdown start time, for all three models. The inclusion of nonlinear effects stabilizes the inferred amplitude, which otherwise would be biased. \textbf{Right:} Mismatch for the linear, quadratic, and cubic model as a function of the ringdown start time. The inclusion of second-order effects improves more than one order of magnitude compared to the linear model. Adding also cubic effects improves the fit by an extra order of magnitude.}
    \label{fig:evidence}
\end{figure*}

Next, we focus on the simulation with the largest collision velocity, where nonlinear effects are expected to be most significant. We compare the performance of the three models: linear (L), quadratic (Q), and cubic (C). The results are summarized in Fig.~\ref{fig:evidence}.

The left panel shows the best-fit signals together with their residuals for each model. While all three provide an overall good description of the ringdown, the residuals make clear that higher-order corrections systematically improve the fit. This is remarkable given that all models involve the same number of free parameters.

The middle panel shows the extracted amplitude $A$ and phase $\phi$ of the linear QNM as a function of the ringdown start time $t_0$, using each model.

Two points are worth emphasizing: (i) including nonlinear effects stabilizes the inferred values of $A$ and $\phi$ at earlier times, and (ii) neglecting them leads to a biased estimate of the linear QNM parameters. Once nonlinearities are included, a consistent extraction of the linear mode is already possible close to the merger time, $t_0 \sim t_{\rm mg}$.

Finally, the right panel of Fig.~\ref{fig:evidence} shows the mismatch $\mathcal{M}$ defined in eq.~\eqref{eq:mismatch} as a function of $t_0$. Both quadratic and cubic models improve the mismatch by roughly an order of magnitude compared to the linear case. Taken together, these results provide clear evidence of nonlinear effects in black hole ringdown in the large-$D$ limit of general relativity, and show that they are quantitatively captured by perturbation theory beyond linear order. This also provides evidence of cubic-order contributions in black hole ringdown following a merger (see e.g.~\cite{Sberna:2021eui, Redondo-Yuste:2023ipg, Zhu:2024dyl, May:2024rrg, Dyer:2025hdt}). In particular, cubic QNMs are directly excited and significantly improve the modelling of the ringdown signal.

% \begin{figure}[]
%     \centering
%     \includegraphics[width=\columnwidth]{figs/fig_min_mismatch_vs_v.pdf}
%     \caption{Minimum mismatch achieved with the linear, quadratic, and cubic models, as a function of the merger velocity $v$. The nonlinear models improve more than one order of magnitude the fit, with the cubic model outperforming the quadratic one at high merger velocities. }
%     \label{fig:mismatch}
% \end{figure}

Finally, we investigate the dependence on the merger velocity. We expect that nonlinear effects become more pronounced at larger $v$, and Fig.~\ref{fig:qqnms} already indicates this trend. To quantify it, we consider the \emph{minimum} mismatch over the fitting window $t_0 \in [t_{\rm mg} -0.5, t_{\rm mg}+2]$, while keeping the ringdown end time $t_{\rm end}$ fixed. We evaluate this quantity for the linear (L), quadratic (Q), and cubic (C) models across a range of velocities, as shown in the left panel of Fig.~\ref{fig:mismatch}.

At low velocities, quadratic corrections already produce a substantial improvement, reducing the mismatch by nearly two orders of magnitude, while cubic effects are negligible. As the velocity increases, however, the performance of the quadratic model degrades, whereas the cubic model remains consistently accurate. This indicates that third-order contributions become progressively more important with increasing collision velocity. 

Pushing to larger velocities clearly suggests a systematic enhancement of higher-order effects, reflecting increasingly nonlinear regimes of the effective theory. It should nonetheless be kept in mind that velocities in the large-$D$ description always remain non-relativistic, so the apparent trend towards a formal $v\to\infty$ limit cannot be taken literally in a relativistic sense. Still, it offers useful qualitative guidance on how nonlinear effects would become increasingly important in more relativistic, finite-$D$ mergers.

\subsection{Lazarus Evolution}

The simplicity of black hole dynamics in the large-$D$ limit allows us to implement a complete version of the \emph{Lazarus} program~\cite{Baker:2000zh, Baker:2001nu, Baker:2001sf, Campanelli:2005ia}. The original idea was to combine nonlinear numerical relativity for the inspiral with a linearized close-limit approximation for the merger–ringdown phase~\cite{Price:1994pm, Gleiser:1996yc, Khanna:1999mh}, and to assess the role of nonlinearities by comparing the resulting waveform with full nonlinear evolutions.

In the present setting, and particularly for head-on collisions, this construction becomes especially simple. Instead of evolving the full nonlinear system throughout, we introduce a switch time $t_{\rm sw}$ at which the evolution is matched onto a linearized regime. For $t>t_{\rm sw}$, we therefore evolve the linearization of the effective equations~\eqref{eq:effective_eqs} around the remnant black hole background~\eqref{eq:remnant},
\begin{equation}
    \begin{aligned}
        \partial_t\delta m-\partial^2_{zz}\delta m =& -\partial_z \delta p \, , \\
        \partial_t \delta p -\partial^2_{zz}\delta p =& \partial_z \delta m +2z(\delta m+\partial_z \delta p)+z^2 \partial_z \delta m - 2\delta p \, . 
    \end{aligned}
\end{equation}
with Dirichlet boundary conditions imposed at the outer edges of the numerical grid, and initial data specified by 
\begin{equation}
    \begin{aligned}
        \delta m(t=t_{\rm sw}) =& m(t_{\rm sw},z) - m_{\rm rem}(z) \, , \\
        \delta p(t=t_{\rm sw}) =& p(t_{\rm sw},z) - p_{\rm rem}(z) \, .
    \end{aligned}
\end{equation}
\begin{figure}[t!]
    \centering
    \includegraphics[width=\columnwidth]{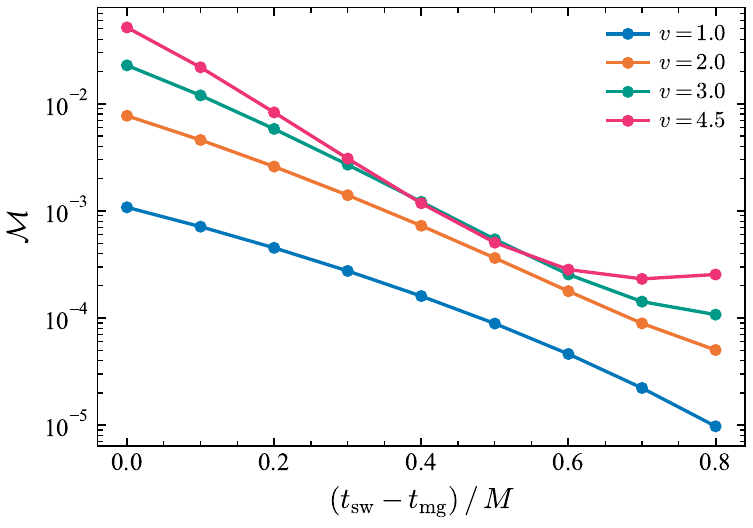}
    \caption{Mismatch between the Lazarus evolution(where linear perturbation theory is applied after a switch time $t_{\rm sw}$) and the full nonlinear solution, as a function of $t_{\rm sw}$ and for different merger velocities, as indicated in the legend.}
    \label{fig:lazarus}
\end{figure}

We then compute the mismatch $\mathcal{M}$ defined in \eqref{eq:mismatch} between the $\ell=2$ mode extracted from the full nonlinear evolution and the corresponding Lazarus evolution with switch time $t_{\rm sw}$. The results are shown in Fig.~\ref{fig:lazarus}. As expected, delaying the switch time systematically improves the agreement. This is consistent with the perturbative picture, in which nonlinear contributions to the $\ell=2$ mode are purely higher-order QNMs and therefore decay exponentially in time. The results also reinforce the conclusion that nonlinear effects become more important as the merger velocity increases, in agreement with the fits of the full waveform shown in Fig.~\ref{fig:mismatch}.

A natural refinement of the Lazarus construction would include second-order perturbative effects, which should further improve the agreement, particularly at larger velocities. Beyond its technical implementation, this comparison provides a direct and transparent measure of the size of nonlinear contributions in the ringdown. A corresponding analysis in $D=4$, comparing Lazarus-type evolutions with high-precision numerical-relativity waveforms~\cite{Ma:2023qjn, Scheel:2025jct}, would be valuable for quantifying nonlinear effects across different harmonics and binary parameters.

%%%%%%%%%%%%%%%%%%%%
\section{Discussion}\label{sec:Discussion}
%%%%%%%%%%%%%%%%%%%%

In this work we have characterised nonlinear effects in black hole ringdown in the large-$D$ limit of general relativity. This provides a first-principles framework in which several questions in black hole perturbation theory become analytically tractable. In particular, we have developed a systematic construction of nonlinear QNM amplitudes at arbitrary order and explicitly computed the leading cubic contributions. Comparing these predictions with numerical simulations of head-on, equal-mass mergers, we have shown that nonlinear QNMs are indeed excited during the merger–ringdown phase. Including quadratic and cubic effects improves the modelling of the relaxation dynamics by several orders of magnitude, with cubic QNMs becoming particularly relevant at higher collision velocities.

We have also introduced ``Lazarus-type'' hybrid evolutions, in which the system is switched from a nonlinear to a perturbative description at an intermediate time. This provides a quantitative probe of the onset and strength of nonlinearities near the merger, and allows us to track their dependence on the binary parameters in a controlled way.

While we do not expect the large-$D$ limit to provide a quantitatively accurate description of astrophysical black holes in $D=4$, it does offer a valuable qualitative laboratory for exploring nonlinear dynamics that are otherwise difficult to access. In this context, a key result is the role of cubic QNMs, which we have analysed here for the first time in this setting. These modes exhibit a pseudo-resonant enhancement when their real frequency matches that of the linear parent mode, making them potentially important in modelling merger–ringdown waveforms. Our results show that neglecting them can bias the inference of linear QNM amplitudes.

\medskip
\noindent\textit{Outlook.} Several directions remain open. From a theoretical perspective, it would be interesting to clarify the role of nonlinearities in the entropy growth during the merger, which can be evaluated within the effective theory using the entropy current introduced  in~\cite{Andrade:2020ilm}. Reintroducing the coupling to the radiation sector---decoupled at all perturbative orders in $1/D$---seems in principle possible, although technically involved, using the effective stress-tensor of~\cite{Bhattacharyya:2016nhn}. Extensions to asymmetric and spinning mergers, as well as a systematic study of the convergence properties of the nonlinear QNM expansion, are particularly promising directions that we intend to explore in future work.

%%%%%%%%%%%%%%%%%%%%%%%%%%%%%%
\section*{Acknowledgements}
%%%%%%%%%%%%%%%%%%%%%%%%%%%%%%
We thank Emanuele Berti for valuable discussions. 
D.~P.\ and J.~R.-Y.\ are supported by NSF Grants No.~AST-2307146, No.~PHY-2513337, No.~PHY-090003, and No.~PHY-20043, by NASA Grant No.~21-ATP21-0010, by John Templeton Foundation Grant No.~62840, by the Simons Foundation [MPS-SIP-00001698, E.B.], by the Simons Foundation International [SFI-MPS-BH-00012593-02], and by Italian Ministry of Foreign Affairs and International Cooperation Grant No.~PGR01167. R.~E. and A.~G.-S. are supported by MICINN grant PID2022-136224NBC22, AGAUR grant 2021 SGR 00872, and the “Unit of Excellence María de Maeztu” grant CEX2024-001451-M funded by MICIU/AEI/10.13039/501100011033.  A.~G.-S. also acknowledges support from the Leverhulme Trust through a Study Abroad Studentship.

\bibliography{ref}

\begin{thebibliography}{70}%
\makeatletter
\providecommand \@ifxundefined [1]{%
 \@ifx{#1\undefined}
}%
\providecommand \@ifnum [1]{%
 \ifnum #1\expandafter \@firstoftwo
 \else \expandafter \@secondoftwo
 \fi
}%
\providecommand \@ifx [1]{%
 \ifx #1\expandafter \@firstoftwo
 \else \expandafter \@secondoftwo
 \fi
}%
\providecommand \natexlab [1]{#1}%
\providecommand \enquote  [1]{``#1''}%
\providecommand \bibnamefont  [1]{#1}%
\providecommand \bibfnamefont [1]{#1}%
\providecommand \citenamefont [1]{#1}%
\providecommand \href@noop [0]{\@secondoftwo}%
\providecommand \href [0]{\begingroup \@sanitize@url \@href}%
\providecommand \@href[1]{\@@startlink{#1}\@@href}%
\providecommand \@@href[1]{\endgroup#1\@@endlink}%
\providecommand \@sanitize@url [0]{\catcode `\\12\catcode `\$12\catcode
  `\&12\catcode `\#12\catcode `\^12\catcode `\_12\catcode `\%12\relax}%
\providecommand \@@startlink[1]{}%
\providecommand \@@endlink[0]{}%
\providecommand \url  [0]{\begingroup\@sanitize@url \@url }%
\providecommand \@url [1]{\endgroup\@href {#1}{\urlprefix }}%
\providecommand \urlprefix  [0]{URL }%
\providecommand \Eprint [0]{\href }%
\providecommand \doibase [0]{http://dx.doi.org/}%
\providecommand \selectlanguage [0]{\@gobble}%
\providecommand \bibinfo  [0]{\@secondoftwo}%
\providecommand \bibfield  [0]{\@secondoftwo}%
\providecommand \translation [1]{[#1]}%
\providecommand \BibitemOpen [0]{}%
\providecommand \bibitemStop [0]{}%
\providecommand \bibitemNoStop [0]{.\EOS\space}%
\providecommand \EOS [0]{\spacefactor3000\relax}%
\providecommand \BibitemShut  [1]{\csname bibitem#1\endcsname}%
\let\auto@bib@innerbib\@empty
%</preamble>
\bibitem [{\citenamefont {London}\ \emph {et~al.}(2014)\citenamefont {London},
  \citenamefont {Shoemaker},\ and\ \citenamefont {Healy}}]{London:2014cma}%
  \BibitemOpen
  \bibfield  {author} {\bibinfo {author} {\bibfnamefont {Lionel}\ \bibnamefont
  {London}}, \bibinfo {author} {\bibfnamefont {Deirdre}\ \bibnamefont
  {Shoemaker}}, \ and\ \bibinfo {author} {\bibfnamefont {James}\ \bibnamefont
  {Healy}},\ }\bibfield  {title} {\enquote {\bibinfo {title} {{Modeling
  ringdown: Beyond the fundamental quasinormal modes}},}\ }\href {\doibase
  10.1103/PhysRevD.90.124032} {\bibfield  {journal} {\bibinfo  {journal} {Phys.
  Rev. D}\ }\textbf {\bibinfo {volume} {90}},\ \bibinfo {pages} {124032}
  (\bibinfo {year} {2014})},\ \bibinfo {note} {[Erratum: Phys.Rev.D 94, 069902
  (2016)]},\ \Eprint {http://arxiv.org/abs/1404.3197} {arXiv:1404.3197 [gr-qc]}
  \BibitemShut {NoStop}%
\bibitem [{\citenamefont {Cheung}\ \emph {et~al.}(2023)\citenamefont {Cheung}
  \emph {et~al.}}]{Cheung:2022rbm}%
  \BibitemOpen
  \bibfield  {author} {\bibinfo {author} {\bibfnamefont {Mark Ho-Yeuk}\
  \bibnamefont {Cheung}} \emph {et~al.},\ }\bibfield  {title} {\enquote
  {\bibinfo {title} {{Nonlinear Effects in Black Hole Ringdown}},}\ }\href
  {\doibase 10.1103/PhysRevLett.130.081401} {\bibfield  {journal} {\bibinfo
  {journal} {Phys. Rev. Lett.}\ }\textbf {\bibinfo {volume} {130}},\ \bibinfo
  {pages} {081401} (\bibinfo {year} {2023})},\ \Eprint
  {http://arxiv.org/abs/2208.07374} {arXiv:2208.07374 [gr-qc]} \BibitemShut
  {NoStop}%
\bibitem [{\citenamefont {Mitman}\ \emph {et~al.}(2023)\citenamefont {Mitman}
  \emph {et~al.}}]{Mitman:2022qdl}%
  \BibitemOpen
  \bibfield  {author} {\bibinfo {author} {\bibfnamefont {Keefe}\ \bibnamefont
  {Mitman}} \emph {et~al.},\ }\bibfield  {title} {\enquote {\bibinfo {title}
  {{Nonlinearities in Black Hole Ringdowns}},}\ }\href {\doibase
  10.1103/PhysRevLett.130.081402} {\bibfield  {journal} {\bibinfo  {journal}
  {Phys. Rev. Lett.}\ }\textbf {\bibinfo {volume} {130}},\ \bibinfo {pages}
  {081402} (\bibinfo {year} {2023})},\ \Eprint
  {http://arxiv.org/abs/2208.07380} {arXiv:2208.07380 [gr-qc]} \BibitemShut
  {NoStop}%
\bibitem [{\citenamefont {Zlochower}\ \emph {et~al.}(2003)\citenamefont
  {Zlochower}, \citenamefont {Gomez}, \citenamefont {Husa}, \citenamefont
  {Lehner},\ and\ \citenamefont {Winicour}}]{Zlochower:2003yh}%
  \BibitemOpen
  \bibfield  {author} {\bibinfo {author} {\bibfnamefont {Yosef}\ \bibnamefont
  {Zlochower}}, \bibinfo {author} {\bibfnamefont {Roberto}\ \bibnamefont
  {Gomez}}, \bibinfo {author} {\bibfnamefont {Sascha}\ \bibnamefont {Husa}},
  \bibinfo {author} {\bibfnamefont {Luis}\ \bibnamefont {Lehner}}, \ and\
  \bibinfo {author} {\bibfnamefont {Jeffrey}\ \bibnamefont {Winicour}},\
  }\bibfield  {title} {\enquote {\bibinfo {title} {{Mode coupling in the
  nonlinear response of black holes}},}\ }\href {\doibase
  10.1103/PhysRevD.68.084014} {\bibfield  {journal} {\bibinfo  {journal} {Phys.
  Rev. D}\ }\textbf {\bibinfo {volume} {68}},\ \bibinfo {pages} {084014}
  (\bibinfo {year} {2003})},\ \Eprint {http://arxiv.org/abs/gr-qc/0306098}
  {arXiv:gr-qc/0306098} \BibitemShut {NoStop}%
\bibitem [{\citenamefont {Yi}\ \emph {et~al.}(2024)\citenamefont {Yi},
  \citenamefont {Kuntz}, \citenamefont {Barausse}, \citenamefont {Berti},
  \citenamefont {Cheung}, \citenamefont {Kritos},\ and\ \citenamefont
  {Maselli}}]{Yi:2024elj}%
  \BibitemOpen
  \bibfield  {author} {\bibinfo {author} {\bibfnamefont {Sophia}\ \bibnamefont
  {Yi}}, \bibinfo {author} {\bibfnamefont {Adrien}\ \bibnamefont {Kuntz}},
  \bibinfo {author} {\bibfnamefont {Enrico}\ \bibnamefont {Barausse}}, \bibinfo
  {author} {\bibfnamefont {Emanuele}\ \bibnamefont {Berti}}, \bibinfo {author}
  {\bibfnamefont {Mark Ho-Yeuk}\ \bibnamefont {Cheung}}, \bibinfo {author}
  {\bibfnamefont {Konstantinos}\ \bibnamefont {Kritos}}, \ and\ \bibinfo
  {author} {\bibfnamefont {Andrea}\ \bibnamefont {Maselli}},\ }\bibfield
  {title} {\enquote {\bibinfo {title} {{Nonlinear quasinormal mode
  detectability with next-generation gravitational wave detectors}},}\ }\href
  {\doibase 10.1103/PhysRevD.109.124029} {\bibfield  {journal} {\bibinfo
  {journal} {Phys. Rev. D}\ }\textbf {\bibinfo {volume} {109}},\ \bibinfo
  {pages} {124029} (\bibinfo {year} {2024})},\ \Eprint
  {http://arxiv.org/abs/2403.09767} {arXiv:2403.09767 [gr-qc]} \BibitemShut
  {NoStop}%
\bibitem [{\citenamefont {Lagos}\ \emph {et~al.}(2025)\citenamefont {Lagos},
  \citenamefont {Andrade}, \citenamefont {Rafecas-Ventosa},\ and\ \citenamefont
  {Hui}}]{Lagos:2024ekd}%
  \BibitemOpen
  \bibfield  {author} {\bibinfo {author} {\bibfnamefont {Macarena}\
  \bibnamefont {Lagos}}, \bibinfo {author} {\bibfnamefont {Tom{\'a}s}\
  \bibnamefont {Andrade}}, \bibinfo {author} {\bibfnamefont {Jordi}\
  \bibnamefont {Rafecas-Ventosa}}, \ and\ \bibinfo {author} {\bibfnamefont
  {Lam}\ \bibnamefont {Hui}},\ }\bibfield  {title} {\enquote {\bibinfo {title}
  {{Black hole spectroscopy with nonlinear quasinormal modes}},}\ }\href
  {\doibase 10.1103/PhysRevD.111.024018} {\bibfield  {journal} {\bibinfo
  {journal} {Phys. Rev. D}\ }\textbf {\bibinfo {volume} {111}},\ \bibinfo
  {pages} {024018} (\bibinfo {year} {2025})},\ \Eprint
  {http://arxiv.org/abs/2411.02264} {arXiv:2411.02264 [gr-qc]} \BibitemShut
  {NoStop}%
\bibitem [{\citenamefont {Wang}\ \emph {et~al.}(2026)\citenamefont {Wang},
  \citenamefont {Ma}, \citenamefont {Khera},\ and\ \citenamefont
  {Yang}}]{Wang:2026rev}%
  \BibitemOpen
  \bibfield  {author} {\bibinfo {author} {\bibfnamefont {Yi-Fan}\ \bibnamefont
  {Wang}}, \bibinfo {author} {\bibfnamefont {Sizheng}\ \bibnamefont {Ma}},
  \bibinfo {author} {\bibfnamefont {Neev}\ \bibnamefont {Khera}}, \ and\
  \bibinfo {author} {\bibfnamefont {Huan}\ \bibnamefont {Yang}},\ }\bibfield
  {title} {\enquote {\bibinfo {title} {{A nonlinear voice from GW250114
  ringdown}},}\ }\href@noop {} {\  (\bibinfo {year} {2026})},\ \Eprint
  {http://arxiv.org/abs/2601.05734} {arXiv:2601.05734 [gr-qc]} \BibitemShut
  {NoStop}%
\bibitem [{\citenamefont {Gleiser}\ \emph
  {et~al.}(1996{\natexlab{a}})\citenamefont {Gleiser}, \citenamefont {Nicasio},
  \citenamefont {Price},\ and\ \citenamefont {Pullin}}]{Gleiser:1995gx}%
  \BibitemOpen
  \bibfield  {author} {\bibinfo {author} {\bibfnamefont {Reinaldo~J.}\
  \bibnamefont {Gleiser}}, \bibinfo {author} {\bibfnamefont {Carlos~O.}\
  \bibnamefont {Nicasio}}, \bibinfo {author} {\bibfnamefont {Richard~H.}\
  \bibnamefont {Price}}, \ and\ \bibinfo {author} {\bibfnamefont {Jorge}\
  \bibnamefont {Pullin}},\ }\bibfield  {title} {\enquote {\bibinfo {title}
  {{Second order perturbations of a Schwarzschild black hole}},}\ }\href
  {\doibase 10.1088/0264-9381/13/10/001} {\bibfield  {journal} {\bibinfo
  {journal} {Class. Quant. Grav.}\ }\textbf {\bibinfo {volume} {13}},\ \bibinfo
  {pages} {L117--L124} (\bibinfo {year} {1996}{\natexlab{a}})},\ \Eprint
  {http://arxiv.org/abs/gr-qc/9510049} {arXiv:gr-qc/9510049} \BibitemShut
  {NoStop}%
\bibitem [{\citenamefont {Gleiser}\ \emph
  {et~al.}(1996{\natexlab{b}})\citenamefont {Gleiser}, \citenamefont {Nicasio},
  \citenamefont {Price},\ and\ \citenamefont {Pullin}}]{Gleiser:1996yc}%
  \BibitemOpen
  \bibfield  {author} {\bibinfo {author} {\bibfnamefont {Reinaldo~J.}\
  \bibnamefont {Gleiser}}, \bibinfo {author} {\bibfnamefont {Carlos~O.}\
  \bibnamefont {Nicasio}}, \bibinfo {author} {\bibfnamefont {Richard~H.}\
  \bibnamefont {Price}}, \ and\ \bibinfo {author} {\bibfnamefont {Jorge}\
  \bibnamefont {Pullin}},\ }\bibfield  {title} {\enquote {\bibinfo {title}
  {{Colliding black holes: How far can the close approximation go?}}}\ }\href
  {\doibase 10.1103/PhysRevLett.77.4483} {\bibfield  {journal} {\bibinfo
  {journal} {Phys. Rev. Lett.}\ }\textbf {\bibinfo {volume} {77}},\ \bibinfo
  {pages} {4483--4486} (\bibinfo {year} {1996}{\natexlab{b}})},\ \Eprint
  {http://arxiv.org/abs/gr-qc/9609022} {arXiv:gr-qc/9609022} \BibitemShut
  {NoStop}%
\bibitem [{\citenamefont {Garat}\ and\ \citenamefont
  {Price}(2000)}]{Garat:1999vr}%
  \BibitemOpen
  \bibfield  {author} {\bibinfo {author} {\bibfnamefont {Alcides}\ \bibnamefont
  {Garat}}\ and\ \bibinfo {author} {\bibfnamefont {Richard~H.}\ \bibnamefont
  {Price}},\ }\bibfield  {title} {\enquote {\bibinfo {title} {{Gauge invariant
  formalism for second order perturbations of Schwarzschild space-times}},}\
  }\href {\doibase 10.1103/PhysRevD.61.044006} {\bibfield  {journal} {\bibinfo
  {journal} {Phys. Rev. D}\ }\textbf {\bibinfo {volume} {61}},\ \bibinfo
  {pages} {044006} (\bibinfo {year} {2000})},\ \Eprint
  {http://arxiv.org/abs/gr-qc/9909005} {arXiv:gr-qc/9909005} \BibitemShut
  {NoStop}%
\bibitem [{\citenamefont {Campanelli}\ and\ \citenamefont
  {Lousto}(1999)}]{Campanelli:1998jv}%
  \BibitemOpen
  \bibfield  {author} {\bibinfo {author} {\bibfnamefont {Manuela}\ \bibnamefont
  {Campanelli}}\ and\ \bibinfo {author} {\bibfnamefont {Carlos~O.}\
  \bibnamefont {Lousto}},\ }\bibfield  {title} {\enquote {\bibinfo {title}
  {{Second order gauge invariant gravitational perturbations of a Kerr black
  hole}},}\ }\href {\doibase 10.1103/PhysRevD.59.124022} {\bibfield  {journal}
  {\bibinfo  {journal} {Phys. Rev. D}\ }\textbf {\bibinfo {volume} {59}},\
  \bibinfo {pages} {124022} (\bibinfo {year} {1999})},\ \Eprint
  {http://arxiv.org/abs/gr-qc/9811019} {arXiv:gr-qc/9811019} \BibitemShut
  {NoStop}%
\bibitem [{\citenamefont {Brizuela}\ \emph {et~al.}(2006)\citenamefont
  {Brizuela}, \citenamefont {Martin-Garcia},\ and\ \citenamefont
  {Mena~Marugan}}]{Brizuela:2006ne}%
  \BibitemOpen
  \bibfield  {author} {\bibinfo {author} {\bibfnamefont {David}\ \bibnamefont
  {Brizuela}}, \bibinfo {author} {\bibfnamefont {Jose~M.}\ \bibnamefont
  {Martin-Garcia}}, \ and\ \bibinfo {author} {\bibfnamefont {Guillermo~A.}\
  \bibnamefont {Mena~Marugan}},\ }\bibfield  {title} {\enquote {\bibinfo
  {title} {{Second and higher-order perturbations of a spherical spacetime}},}\
  }\href {\doibase 10.1103/PhysRevD.74.044039} {\bibfield  {journal} {\bibinfo
  {journal} {Phys. Rev. D}\ }\textbf {\bibinfo {volume} {74}},\ \bibinfo
  {pages} {044039} (\bibinfo {year} {2006})},\ \Eprint
  {http://arxiv.org/abs/gr-qc/0607025} {arXiv:gr-qc/0607025} \BibitemShut
  {NoStop}%
\bibitem [{\citenamefont {Brizuela}\ \emph {et~al.}(2007)\citenamefont
  {Brizuela}, \citenamefont {Martin-Garcia},\ and\ \citenamefont
  {Marugan}}]{Brizuela:2007zza}%
  \BibitemOpen
  \bibfield  {author} {\bibinfo {author} {\bibfnamefont {David}\ \bibnamefont
  {Brizuela}}, \bibinfo {author} {\bibfnamefont {Jose~M.}\ \bibnamefont
  {Martin-Garcia}}, \ and\ \bibinfo {author} {\bibfnamefont {Guillermo
  A.~Mena}\ \bibnamefont {Marugan}},\ }\bibfield  {title} {\enquote {\bibinfo
  {title} {{High-order gauge-invariant perturbations of a spherical
  spacetime}},}\ }\href {\doibase 10.1103/PhysRevD.76.024004} {\bibfield
  {journal} {\bibinfo  {journal} {Phys. Rev. D}\ }\textbf {\bibinfo {volume}
  {76}},\ \bibinfo {pages} {024004} (\bibinfo {year} {2007})},\ \Eprint
  {http://arxiv.org/abs/gr-qc/0703069} {arXiv:gr-qc/0703069} \BibitemShut
  {NoStop}%
\bibitem [{\citenamefont {Brizuela}\ \emph {et~al.}(2009)\citenamefont
  {Brizuela}, \citenamefont {Martin-Garcia},\ and\ \citenamefont
  {Tiglio}}]{Brizuela:2009qd}%
  \BibitemOpen
  \bibfield  {author} {\bibinfo {author} {\bibfnamefont {David}\ \bibnamefont
  {Brizuela}}, \bibinfo {author} {\bibfnamefont {Jose~M.}\ \bibnamefont
  {Martin-Garcia}}, \ and\ \bibinfo {author} {\bibfnamefont {Manuel}\
  \bibnamefont {Tiglio}},\ }\bibfield  {title} {\enquote {\bibinfo {title} {{A
  Complete gauge-invariant formalism for arbitrary second-order perturbations
  of a Schwarzschild black hole}},}\ }\href {\doibase
  10.1103/PhysRevD.80.024021} {\bibfield  {journal} {\bibinfo  {journal} {Phys.
  Rev. D}\ }\textbf {\bibinfo {volume} {80}},\ \bibinfo {pages} {024021}
  (\bibinfo {year} {2009})},\ \Eprint {http://arxiv.org/abs/0903.1134}
  {arXiv:0903.1134 [gr-qc]} \BibitemShut {NoStop}%
\bibitem [{\citenamefont {Pazos}\ \emph {et~al.}(2010)\citenamefont {Pazos},
  \citenamefont {Brizuela}, \citenamefont {Martin-Garcia},\ and\ \citenamefont
  {Tiglio}}]{Pazos:2010xf}%
  \BibitemOpen
  \bibfield  {author} {\bibinfo {author} {\bibfnamefont {Enrique}\ \bibnamefont
  {Pazos}}, \bibinfo {author} {\bibfnamefont {David}\ \bibnamefont {Brizuela}},
  \bibinfo {author} {\bibfnamefont {Jose~M.}\ \bibnamefont {Martin-Garcia}}, \
  and\ \bibinfo {author} {\bibfnamefont {Manuel}\ \bibnamefont {Tiglio}},\
  }\bibfield  {title} {\enquote {\bibinfo {title} {{Mode coupling of
  Schwarzschild perturbations: Ringdown frequencies}},}\ }\href {\doibase
  10.1103/PhysRevD.82.104028} {\bibfield  {journal} {\bibinfo  {journal} {Phys.
  Rev. D}\ }\textbf {\bibinfo {volume} {82}},\ \bibinfo {pages} {104028}
  (\bibinfo {year} {2010})},\ \Eprint {http://arxiv.org/abs/1009.4665}
  {arXiv:1009.4665 [gr-qc]} \BibitemShut {NoStop}%
\bibitem [{\citenamefont {Ioka}\ and\ \citenamefont
  {Nakano}(2007)}]{Ioka:2007ak}%
  \BibitemOpen
  \bibfield  {author} {\bibinfo {author} {\bibfnamefont {Kunihito}\
  \bibnamefont {Ioka}}\ and\ \bibinfo {author} {\bibfnamefont {Hiroyuki}\
  \bibnamefont {Nakano}},\ }\bibfield  {title} {\enquote {\bibinfo {title}
  {{Second and higher-order quasi-normal modes in binary black hole
  mergers}},}\ }\href {\doibase 10.1103/PhysRevD.76.061503} {\bibfield
  {journal} {\bibinfo  {journal} {Phys. Rev. D}\ }\textbf {\bibinfo {volume}
  {76}},\ \bibinfo {pages} {061503} (\bibinfo {year} {2007})},\ \Eprint
  {http://arxiv.org/abs/0704.3467} {arXiv:0704.3467 [astro-ph]} \BibitemShut
  {NoStop}%
\bibitem [{\citenamefont {Nakano}\ and\ \citenamefont
  {Ioka}(2007)}]{Nakano:2007cj}%
  \BibitemOpen
  \bibfield  {author} {\bibinfo {author} {\bibfnamefont {Hiroyuki}\
  \bibnamefont {Nakano}}\ and\ \bibinfo {author} {\bibfnamefont {Kunihito}\
  \bibnamefont {Ioka}},\ }\bibfield  {title} {\enquote {\bibinfo {title}
  {{Second Order Quasi-Normal Mode of the Schwarzschild Black Hole}},}\ }\href
  {\doibase 10.1103/PhysRevD.76.084007} {\bibfield  {journal} {\bibinfo
  {journal} {Phys. Rev. D}\ }\textbf {\bibinfo {volume} {76}},\ \bibinfo
  {pages} {084007} (\bibinfo {year} {2007})},\ \Eprint
  {http://arxiv.org/abs/0708.0450} {arXiv:0708.0450 [gr-qc]} \BibitemShut
  {NoStop}%
\bibitem [{\citenamefont {Lagos}\ and\ \citenamefont
  {Hui}(2023)}]{Lagos:2022otp}%
  \BibitemOpen
  \bibfield  {author} {\bibinfo {author} {\bibfnamefont {Macarena}\
  \bibnamefont {Lagos}}\ and\ \bibinfo {author} {\bibfnamefont {Lam}\
  \bibnamefont {Hui}},\ }\bibfield  {title} {\enquote {\bibinfo {title}
  {{Generation and propagation of nonlinear quasinormal modes of a
  Schwarzschild black hole}},}\ }\href {\doibase 10.1103/PhysRevD.107.044040}
  {\bibfield  {journal} {\bibinfo  {journal} {Phys. Rev. D}\ }\textbf {\bibinfo
  {volume} {107}},\ \bibinfo {pages} {044040} (\bibinfo {year} {2023})},\
  \Eprint {http://arxiv.org/abs/2208.07379} {arXiv:2208.07379 [gr-qc]}
  \BibitemShut {NoStop}%
\bibitem [{\citenamefont {Bucciotti}\ \emph {et~al.}(2023)\citenamefont
  {Bucciotti}, \citenamefont {Kuntz}, \citenamefont {Serra},\ and\
  \citenamefont {Trincherini}}]{Bucciotti:2023ets}%
  \BibitemOpen
  \bibfield  {author} {\bibinfo {author} {\bibfnamefont {Bruno}\ \bibnamefont
  {Bucciotti}}, \bibinfo {author} {\bibfnamefont {Adrien}\ \bibnamefont
  {Kuntz}}, \bibinfo {author} {\bibfnamefont {Francesco}\ \bibnamefont
  {Serra}}, \ and\ \bibinfo {author} {\bibfnamefont {Enrico}\ \bibnamefont
  {Trincherini}},\ }\bibfield  {title} {\enquote {\bibinfo {title} {{Nonlinear
  quasi-normal modes: uniform approximation}},}\ }\href {\doibase
  10.1007/JHEP12(2023)048} {\bibfield  {journal} {\bibinfo  {journal} {JHEP}\
  }\textbf {\bibinfo {volume} {12}},\ \bibinfo {pages} {048} (\bibinfo {year}
  {2023})},\ \Eprint {http://arxiv.org/abs/2309.08501} {arXiv:2309.08501
  [hep-th]} \BibitemShut {NoStop}%
\bibitem [{\citenamefont {Khera}\ \emph {et~al.}(2023)\citenamefont {Khera},
  \citenamefont {Ribes~Metidieri}, \citenamefont {Bonga}, \citenamefont
  {Jim{\'e}nez~Forteza}, \citenamefont {Krishnan}, \citenamefont {Poisson},
  \citenamefont {Pook-Kolb}, \citenamefont {Schnetter},\ and\ \citenamefont
  {Yang}}]{Khera:2023oyf}%
  \BibitemOpen
  \bibfield  {author} {\bibinfo {author} {\bibfnamefont {Neev}\ \bibnamefont
  {Khera}}, \bibinfo {author} {\bibfnamefont {Ariadna}\ \bibnamefont
  {Ribes~Metidieri}}, \bibinfo {author} {\bibfnamefont {B{\'e}atrice}\
  \bibnamefont {Bonga}}, \bibinfo {author} {\bibfnamefont {Xisco}\ \bibnamefont
  {Jim{\'e}nez~Forteza}}, \bibinfo {author} {\bibfnamefont {Badri}\
  \bibnamefont {Krishnan}}, \bibinfo {author} {\bibfnamefont {Eric}\
  \bibnamefont {Poisson}}, \bibinfo {author} {\bibfnamefont {Daniel}\
  \bibnamefont {Pook-Kolb}}, \bibinfo {author} {\bibfnamefont {Erik}\
  \bibnamefont {Schnetter}}, \ and\ \bibinfo {author} {\bibfnamefont {Huan}\
  \bibnamefont {Yang}},\ }\bibfield  {title} {\enquote {\bibinfo {title}
  {{Nonlinear Ringdown at the Black Hole Horizon}},}\ }\href {\doibase
  10.1103/PhysRevLett.131.231401} {\bibfield  {journal} {\bibinfo  {journal}
  {Phys. Rev. Lett.}\ }\textbf {\bibinfo {volume} {131}},\ \bibinfo {pages}
  {231401} (\bibinfo {year} {2023})},\ \Eprint
  {http://arxiv.org/abs/2306.11142} {arXiv:2306.11142 [gr-qc]} \BibitemShut
  {NoStop}%
\bibitem [{\citenamefont {Bucciotti}\ \emph
  {et~al.}(2024{\natexlab{a}})\citenamefont {Bucciotti}, \citenamefont
  {Juliano}, \citenamefont {Kuntz},\ and\ \citenamefont
  {Trincherini}}]{Bucciotti:2024zyp}%
  \BibitemOpen
  \bibfield  {author} {\bibinfo {author} {\bibfnamefont {Bruno}\ \bibnamefont
  {Bucciotti}}, \bibinfo {author} {\bibfnamefont {Leonardo}\ \bibnamefont
  {Juliano}}, \bibinfo {author} {\bibfnamefont {Adrien}\ \bibnamefont {Kuntz}},
  \ and\ \bibinfo {author} {\bibfnamefont {Enrico}\ \bibnamefont
  {Trincherini}},\ }\bibfield  {title} {\enquote {\bibinfo {title} {{Quadratic
  Quasi-Normal Modes of a Schwarzschild Black Hole}},}\ }\href@noop {} {\
  (\bibinfo {year} {2024}{\natexlab{a}})},\ \Eprint
  {http://arxiv.org/abs/2405.06012} {arXiv:2405.06012 [gr-qc]} \BibitemShut
  {NoStop}%
\bibitem [{\citenamefont {Bucciotti}\ \emph
  {et~al.}(2024{\natexlab{b}})\citenamefont {Bucciotti}, \citenamefont
  {Juliano}, \citenamefont {Kuntz},\ and\ \citenamefont
  {Trincherini}}]{Bucciotti:2024jrv}%
  \BibitemOpen
  \bibfield  {author} {\bibinfo {author} {\bibfnamefont {Bruno}\ \bibnamefont
  {Bucciotti}}, \bibinfo {author} {\bibfnamefont {Leonardo}\ \bibnamefont
  {Juliano}}, \bibinfo {author} {\bibfnamefont {Adrien}\ \bibnamefont {Kuntz}},
  \ and\ \bibinfo {author} {\bibfnamefont {Enrico}\ \bibnamefont
  {Trincherini}},\ }\bibfield  {title} {\enquote {\bibinfo {title} {{Amplitudes
  and Polarizations of Quadratic Quasi-Normal Modes for a Schwarzschild Black
  Hole}},}\ }\href@noop {} {\  (\bibinfo {year} {2024}{\natexlab{b}})},\
  \Eprint {http://arxiv.org/abs/2406.14611} {arXiv:2406.14611 [hep-th]}
  \BibitemShut {NoStop}%
\bibitem [{\citenamefont {Ben~Achour}\ and\ \citenamefont
  {Roussille}(2024)}]{BenAchour:2024skv}%
  \BibitemOpen
  \bibfield  {author} {\bibinfo {author} {\bibfnamefont {Jibril}\ \bibnamefont
  {Ben~Achour}}\ and\ \bibinfo {author} {\bibfnamefont {Hugo}\ \bibnamefont
  {Roussille}},\ }\bibfield  {title} {\enquote {\bibinfo {title} {{Quadratic
  perturbations of the Schwarzschild black hole: the algebraically special
  sector}},}\ }\href {\doibase 10.1088/1475-7516/2024/07/085} {\bibfield
  {journal} {\bibinfo  {journal} {JCAP}\ }\textbf {\bibinfo {volume} {07}},\
  \bibinfo {pages} {085} (\bibinfo {year} {2024})},\ \Eprint
  {http://arxiv.org/abs/2406.08159} {arXiv:2406.08159 [gr-qc]} \BibitemShut
  {NoStop}%
\bibitem [{\citenamefont {Ma}\ and\ \citenamefont {Yang}(2024)}]{Ma:2024qcv}%
  \BibitemOpen
  \bibfield  {author} {\bibinfo {author} {\bibfnamefont {Sizheng}\ \bibnamefont
  {Ma}}\ and\ \bibinfo {author} {\bibfnamefont {Huan}\ \bibnamefont {Yang}},\
  }\bibfield  {title} {\enquote {\bibinfo {title} {{Excitation of quadratic
  quasinormal modes for Kerr black holes}},}\ }\href {\doibase
  10.1103/PhysRevD.109.104070} {\bibfield  {journal} {\bibinfo  {journal}
  {Phys. Rev. D}\ }\textbf {\bibinfo {volume} {109}},\ \bibinfo {pages}
  {104070} (\bibinfo {year} {2024})},\ \Eprint
  {http://arxiv.org/abs/2401.15516} {arXiv:2401.15516 [gr-qc]} \BibitemShut
  {NoStop}%
\bibitem [{\citenamefont {Bourg}\ \emph {et~al.}(2024)\citenamefont {Bourg},
  \citenamefont {Panosso~Macedo}, \citenamefont {Spiers}, \citenamefont
  {Leather}, \citenamefont {Bonga},\ and\ \citenamefont
  {Pound}}]{Bourg:2024jme}%
  \BibitemOpen
  \bibfield  {author} {\bibinfo {author} {\bibfnamefont {Patrick}\ \bibnamefont
  {Bourg}}, \bibinfo {author} {\bibfnamefont {Rodrigo}\ \bibnamefont
  {Panosso~Macedo}}, \bibinfo {author} {\bibfnamefont {Andrew}\ \bibnamefont
  {Spiers}}, \bibinfo {author} {\bibfnamefont {Benjamin}\ \bibnamefont
  {Leather}}, \bibinfo {author} {\bibfnamefont {B\'eatrice}\ \bibnamefont
  {Bonga}}, \ and\ \bibinfo {author} {\bibfnamefont {Adam}\ \bibnamefont
  {Pound}},\ }\bibfield  {title} {\enquote {\bibinfo {title} {{Quadratic
  quasi-normal mode dependence on linear mode parity}},}\ }\href@noop {} {\
  (\bibinfo {year} {2024})},\ \Eprint {http://arxiv.org/abs/2405.10270}
  {arXiv:2405.10270 [gr-qc]} \BibitemShut {NoStop}%
\bibitem [{\citenamefont {Singh}\ and\ \citenamefont
  {Suneeta}(2025)}]{Singh:2025xzd}%
  \BibitemOpen
  \bibfield  {author} {\bibinfo {author} {\bibfnamefont {Jasveer}\ \bibnamefont
  {Singh}}\ and\ \bibinfo {author} {\bibfnamefont {Vardarajan}\ \bibnamefont
  {Suneeta}},\ }\bibfield  {title} {\enquote {\bibinfo {title} {{Computing
  nonlinearity ratios using second order black hole perturbation theory}},}\
  }\href@noop {} {\  (\bibinfo {year} {2025})},\ \Eprint
  {http://arxiv.org/abs/2512.00943} {arXiv:2512.00943 [gr-qc]} \BibitemShut
  {NoStop}%
\bibitem [{\citenamefont {Redondo-Yuste}\ \emph
  {et~al.}(2024{\natexlab{a}})\citenamefont {Redondo-Yuste}, \citenamefont
  {Carullo}, \citenamefont {Ripley}, \citenamefont {Berti},\ and\ \citenamefont
  {Cardoso}}]{Redondo-Yuste:2023seq}%
  \BibitemOpen
  \bibfield  {author} {\bibinfo {author} {\bibfnamefont {Jaime}\ \bibnamefont
  {Redondo-Yuste}}, \bibinfo {author} {\bibfnamefont {Gregorio}\ \bibnamefont
  {Carullo}}, \bibinfo {author} {\bibfnamefont {Justin~L.}\ \bibnamefont
  {Ripley}}, \bibinfo {author} {\bibfnamefont {Emanuele}\ \bibnamefont
  {Berti}}, \ and\ \bibinfo {author} {\bibfnamefont {Vitor}\ \bibnamefont
  {Cardoso}},\ }\bibfield  {title} {\enquote {\bibinfo {title} {{Spin
  dependence of black hole ringdown nonlinearities}},}\ }\href {\doibase
  10.1103/PhysRevD.109.L101503} {\bibfield  {journal} {\bibinfo  {journal}
  {Phys. Rev. D}\ }\textbf {\bibinfo {volume} {109}},\ \bibinfo {pages}
  {L101503} (\bibinfo {year} {2024}{\natexlab{a}})},\ \Eprint
  {http://arxiv.org/abs/2308.14796} {arXiv:2308.14796 [gr-qc]} \BibitemShut
  {NoStop}%
\bibitem [{\citenamefont {Khera}\ \emph {et~al.}(2025)\citenamefont {Khera},
  \citenamefont {Ma},\ and\ \citenamefont {Yang}}]{Khera:2024bjs}%
  \BibitemOpen
  \bibfield  {author} {\bibinfo {author} {\bibfnamefont {Neev}\ \bibnamefont
  {Khera}}, \bibinfo {author} {\bibfnamefont {Sizheng}\ \bibnamefont {Ma}}, \
  and\ \bibinfo {author} {\bibfnamefont {Huan}\ \bibnamefont {Yang}},\
  }\bibfield  {title} {\enquote {\bibinfo {title} {{Quadratic Mode Couplings in
  Rotating Black Holes and Their Detectability}},}\ }\href {\doibase
  10.1103/PhysRevLett.134.211404} {\bibfield  {journal} {\bibinfo  {journal}
  {Phys. Rev. Lett.}\ }\textbf {\bibinfo {volume} {134}},\ \bibinfo {pages}
  {211404} (\bibinfo {year} {2025})},\ \Eprint
  {http://arxiv.org/abs/2410.14529} {arXiv:2410.14529 [gr-qc]} \BibitemShut
  {NoStop}%
\bibitem [{\citenamefont {Spiers}\ \emph {et~al.}(2023)\citenamefont {Spiers},
  \citenamefont {Pound},\ and\ \citenamefont {Moxon}}]{Spiers:2023cip}%
  \BibitemOpen
  \bibfield  {author} {\bibinfo {author} {\bibfnamefont {Andrew}\ \bibnamefont
  {Spiers}}, \bibinfo {author} {\bibfnamefont {Adam}\ \bibnamefont {Pound}}, \
  and\ \bibinfo {author} {\bibfnamefont {Jordan}\ \bibnamefont {Moxon}},\
  }\bibfield  {title} {\enquote {\bibinfo {title} {{Second-order Teukolsky
  formalism in Kerr spacetime: Formulation and nonlinear source}},}\ }\href
  {\doibase 10.1103/PhysRevD.108.064002} {\bibfield  {journal} {\bibinfo
  {journal} {Phys. Rev. D}\ }\textbf {\bibinfo {volume} {108}},\ \bibinfo
  {pages} {064002} (\bibinfo {year} {2023})},\ \Eprint
  {http://arxiv.org/abs/2305.19332} {arXiv:2305.19332 [gr-qc]} \BibitemShut
  {NoStop}%
\bibitem [{\citenamefont {Spiers}\ \emph {et~al.}(2024)\citenamefont {Spiers},
  \citenamefont {Pound},\ and\ \citenamefont {Wardell}}]{Spiers:2023mor}%
  \BibitemOpen
  \bibfield  {author} {\bibinfo {author} {\bibfnamefont {Andrew}\ \bibnamefont
  {Spiers}}, \bibinfo {author} {\bibfnamefont {Adam}\ \bibnamefont {Pound}}, \
  and\ \bibinfo {author} {\bibfnamefont {Barry}\ \bibnamefont {Wardell}},\
  }\bibfield  {title} {\enquote {\bibinfo {title} {{Second-order perturbations
  of the Schwarzschild spacetime: Practical, covariant, and gauge-invariant
  formalisms}},}\ }\href {\doibase 10.1103/PhysRevD.110.064030} {\bibfield
  {journal} {\bibinfo  {journal} {Phys. Rev. D}\ }\textbf {\bibinfo {volume}
  {110}},\ \bibinfo {pages} {064030} (\bibinfo {year} {2024})},\ \Eprint
  {http://arxiv.org/abs/2306.17847} {arXiv:2306.17847 [gr-qc]} \BibitemShut
  {NoStop}%
\bibitem [{\citenamefont {Bourg}\ \emph {et~al.}(2025)\citenamefont {Bourg},
  \citenamefont {Panosso~Macedo}, \citenamefont {Spiers}, \citenamefont
  {Leather}, \citenamefont {B{\'e}atrice},\ and\ \citenamefont
  {Pound}}]{Bourg:2025lpd}%
  \BibitemOpen
  \bibfield  {author} {\bibinfo {author} {\bibfnamefont {Patrick}\ \bibnamefont
  {Bourg}}, \bibinfo {author} {\bibfnamefont {Rodrigo}\ \bibnamefont
  {Panosso~Macedo}}, \bibinfo {author} {\bibfnamefont {Andrew}\ \bibnamefont
  {Spiers}}, \bibinfo {author} {\bibfnamefont {Benjamin}\ \bibnamefont
  {Leather}}, \bibinfo {author} {\bibfnamefont {Bonga}\ \bibnamefont
  {B{\'e}atrice}}, \ and\ \bibinfo {author} {\bibfnamefont {Adam}\ \bibnamefont
  {Pound}},\ }\bibfield  {title} {\enquote {\bibinfo {title} {{Quadratic
  quasinormal modes at null infinity on a Schwarzschild spacetime}},}\ }\href
  {\doibase 10.1103/fbz4-qsvn} {\bibfield  {journal} {\bibinfo  {journal}
  {Phys. Rev. D}\ }\textbf {\bibinfo {volume} {112}},\ \bibinfo {pages}
  {044049} (\bibinfo {year} {2025})},\ \Eprint
  {http://arxiv.org/abs/2503.07432} {arXiv:2503.07432 [gr-qc]} \BibitemShut
  {NoStop}%
\bibitem [{\citenamefont {Yang}\ \emph {et~al.}(2015)\citenamefont {Yang},
  \citenamefont {Zimmerman},\ and\ \citenamefont {Lehner}}]{Yang:2014tla}%
  \BibitemOpen
  \bibfield  {author} {\bibinfo {author} {\bibfnamefont {Huan}\ \bibnamefont
  {Yang}}, \bibinfo {author} {\bibfnamefont {Aaron}\ \bibnamefont {Zimmerman}},
  \ and\ \bibinfo {author} {\bibfnamefont {Luis}\ \bibnamefont {Lehner}},\
  }\bibfield  {title} {\enquote {\bibinfo {title} {{Turbulent Black Holes}},}\
  }\href {\doibase 10.1103/PhysRevLett.114.081101} {\bibfield  {journal}
  {\bibinfo  {journal} {Phys. Rev. Lett.}\ }\textbf {\bibinfo {volume} {114}},\
  \bibinfo {pages} {081101} (\bibinfo {year} {2015})},\ \Eprint
  {http://arxiv.org/abs/1402.4859} {arXiv:1402.4859 [gr-qc]} \BibitemShut
  {NoStop}%
\bibitem [{\citenamefont {Sberna}\ \emph {et~al.}(2022)\citenamefont {Sberna},
  \citenamefont {Bosch}, \citenamefont {East}, \citenamefont {Green},\ and\
  \citenamefont {Lehner}}]{Sberna:2021eui}%
  \BibitemOpen
  \bibfield  {author} {\bibinfo {author} {\bibfnamefont {Laura}\ \bibnamefont
  {Sberna}}, \bibinfo {author} {\bibfnamefont {Pablo}\ \bibnamefont {Bosch}},
  \bibinfo {author} {\bibfnamefont {William~E.}\ \bibnamefont {East}}, \bibinfo
  {author} {\bibfnamefont {Stephen~R.}\ \bibnamefont {Green}}, \ and\ \bibinfo
  {author} {\bibfnamefont {Luis}\ \bibnamefont {Lehner}},\ }\bibfield  {title}
  {\enquote {\bibinfo {title} {{Nonlinear effects in the black hole ringdown:
  Absorption-induced mode excitation}},}\ }\href {\doibase
  10.1103/PhysRevD.105.064046} {\bibfield  {journal} {\bibinfo  {journal}
  {Phys. Rev. D}\ }\textbf {\bibinfo {volume} {105}},\ \bibinfo {pages}
  {064046} (\bibinfo {year} {2022})},\ \Eprint
  {http://arxiv.org/abs/2112.11168} {arXiv:2112.11168 [gr-qc]} \BibitemShut
  {NoStop}%
\bibitem [{\citenamefont {Redondo-Yuste}\ \emph
  {et~al.}(2024{\natexlab{b}})\citenamefont {Redondo-Yuste}, \citenamefont
  {Pere{\~n}iguez},\ and\ \citenamefont {Cardoso}}]{Redondo-Yuste:2023ipg}%
  \BibitemOpen
  \bibfield  {author} {\bibinfo {author} {\bibfnamefont {Jaime}\ \bibnamefont
  {Redondo-Yuste}}, \bibinfo {author} {\bibfnamefont {David}\ \bibnamefont
  {Pere{\~n}iguez}}, \ and\ \bibinfo {author} {\bibfnamefont {Vitor}\
  \bibnamefont {Cardoso}},\ }\bibfield  {title} {\enquote {\bibinfo {title}
  {{Ringdown of a dynamical spacetime}},}\ }\href {\doibase
  10.1103/PhysRevD.109.044048} {\bibfield  {journal} {\bibinfo  {journal}
  {Phys. Rev. D}\ }\textbf {\bibinfo {volume} {109}},\ \bibinfo {pages}
  {044048} (\bibinfo {year} {2024}{\natexlab{b}})},\ \Eprint
  {http://arxiv.org/abs/2312.04633} {arXiv:2312.04633 [gr-qc]} \BibitemShut
  {NoStop}%
\bibitem [{\citenamefont {Zhu}\ \emph {et~al.}(2024)\citenamefont {Zhu} \emph
  {et~al.}}]{Zhu:2024dyl}%
  \BibitemOpen
  \bibfield  {author} {\bibinfo {author} {\bibfnamefont {Hengrui}\ \bibnamefont
  {Zhu}} \emph {et~al.},\ }\bibfield  {title} {\enquote {\bibinfo {title}
  {{Imprints of changing mass and spin on black hole ringdown}},}\ }\href
  {\doibase 10.1103/PhysRevD.110.124028} {\bibfield  {journal} {\bibinfo
  {journal} {Phys. Rev. D}\ }\textbf {\bibinfo {volume} {110}},\ \bibinfo
  {pages} {124028} (\bibinfo {year} {2024})},\ \Eprint
  {http://arxiv.org/abs/2404.12424} {arXiv:2404.12424 [gr-qc]} \BibitemShut
  {NoStop}%
\bibitem [{\citenamefont {May}\ \emph {et~al.}(2024)\citenamefont {May},
  \citenamefont {Ma}, \citenamefont {Ripley},\ and\ \citenamefont
  {East}}]{May:2024rrg}%
  \BibitemOpen
  \bibfield  {author} {\bibinfo {author} {\bibfnamefont {Taillte}\ \bibnamefont
  {May}}, \bibinfo {author} {\bibfnamefont {Sizheng}\ \bibnamefont {Ma}},
  \bibinfo {author} {\bibfnamefont {Justin~L.}\ \bibnamefont {Ripley}}, \ and\
  \bibinfo {author} {\bibfnamefont {William~E.}\ \bibnamefont {East}},\
  }\bibfield  {title} {\enquote {\bibinfo {title} {{Nonlinear effect of
  absorption on the ringdown of a spinning black hole}},}\ }\href {\doibase
  10.1103/PhysRevD.110.084034} {\bibfield  {journal} {\bibinfo  {journal}
  {Phys. Rev. D}\ }\textbf {\bibinfo {volume} {110}},\ \bibinfo {pages}
  {084034} (\bibinfo {year} {2024})},\ \Eprint
  {http://arxiv.org/abs/2405.18303} {arXiv:2405.18303 [gr-qc]} \BibitemShut
  {NoStop}%
\bibitem [{\citenamefont {Afshordi}\ \emph {et~al.}(2025)\citenamefont
  {Afshordi} \emph {et~al.}}]{LISAConsortiumWaveformWorkingGroup:2023arg}%
  \BibitemOpen
  \bibfield  {author} {\bibinfo {author} {\bibfnamefont {Niayesh}\ \bibnamefont
  {Afshordi}} \emph {et~al.} (\bibinfo {collaboration} {LISA Consortium
  Waveform Working Group}),\ }\bibfield  {title} {\enquote {\bibinfo {title}
  {{Waveform modelling for the Laser Interferometer Space Antenna}},}\ }\href
  {\doibase 10.1007/s41114-025-00056-1} {\bibfield  {journal} {\bibinfo
  {journal} {Living Rev. Rel.}\ }\textbf {\bibinfo {volume} {28}},\ \bibinfo
  {pages} {9} (\bibinfo {year} {2025})},\ \Eprint
  {http://arxiv.org/abs/2311.01300} {arXiv:2311.01300 [gr-qc]} \BibitemShut
  {NoStop}%
\bibitem [{\citenamefont {Emparan}\ \emph {et~al.}(2013)\citenamefont
  {Emparan}, \citenamefont {Suzuki},\ and\ \citenamefont
  {Tanabe}}]{Emparan:2013moa}%
  \BibitemOpen
  \bibfield  {author} {\bibinfo {author} {\bibfnamefont {Roberto}\ \bibnamefont
  {Emparan}}, \bibinfo {author} {\bibfnamefont {Ryotaku}\ \bibnamefont
  {Suzuki}}, \ and\ \bibinfo {author} {\bibfnamefont {Kentaro}\ \bibnamefont
  {Tanabe}},\ }\bibfield  {title} {\enquote {\bibinfo {title} {{The large D
  limit of General Relativity}},}\ }\href {\doibase 10.1007/JHEP06(2013)009}
  {\bibfield  {journal} {\bibinfo  {journal} {JHEP}\ }\textbf {\bibinfo
  {volume} {06}},\ \bibinfo {pages} {009} (\bibinfo {year} {2013})},\ \Eprint
  {http://arxiv.org/abs/1302.6382} {arXiv:1302.6382 [hep-th]} \BibitemShut
  {NoStop}%
\bibitem [{\citenamefont {Emparan}\ \emph
  {et~al.}(2015{\natexlab{a}})\citenamefont {Emparan}, \citenamefont
  {Shiromizu}, \citenamefont {Suzuki}, \citenamefont {Tanabe},\ and\
  \citenamefont {Tanaka}}]{Emparan:2015hwa}%
  \BibitemOpen
  \bibfield  {author} {\bibinfo {author} {\bibfnamefont {Roberto}\ \bibnamefont
  {Emparan}}, \bibinfo {author} {\bibfnamefont {Tetsuya}\ \bibnamefont
  {Shiromizu}}, \bibinfo {author} {\bibfnamefont {Ryotaku}\ \bibnamefont
  {Suzuki}}, \bibinfo {author} {\bibfnamefont {Kentaro}\ \bibnamefont
  {Tanabe}}, \ and\ \bibinfo {author} {\bibfnamefont {Takahiro}\ \bibnamefont
  {Tanaka}},\ }\bibfield  {title} {\enquote {\bibinfo {title} {{Effective
  theory of Black Holes in the 1/D expansion}},}\ }\href {\doibase
  10.1007/JHEP06(2015)159} {\bibfield  {journal} {\bibinfo  {journal} {JHEP}\
  }\textbf {\bibinfo {volume} {06}},\ \bibinfo {pages} {159} (\bibinfo {year}
  {2015}{\natexlab{a}})},\ \Eprint {http://arxiv.org/abs/1504.06489}
  {arXiv:1504.06489 [hep-th]} \BibitemShut {NoStop}%
\bibitem [{\citenamefont {Bhattacharyya}\ \emph
  {et~al.}(2016{\natexlab{a}})\citenamefont {Bhattacharyya}, \citenamefont
  {De}, \citenamefont {Minwalla}, \citenamefont {Mohan},\ and\ \citenamefont
  {Saha}}]{Bhattacharyya:2015dva}%
  \BibitemOpen
  \bibfield  {author} {\bibinfo {author} {\bibfnamefont {Sayantani}\
  \bibnamefont {Bhattacharyya}}, \bibinfo {author} {\bibfnamefont {Anandita}\
  \bibnamefont {De}}, \bibinfo {author} {\bibfnamefont {Shiraz}\ \bibnamefont
  {Minwalla}}, \bibinfo {author} {\bibfnamefont {Ravi}\ \bibnamefont {Mohan}},
  \ and\ \bibinfo {author} {\bibfnamefont {Arunabha}\ \bibnamefont {Saha}},\
  }\bibfield  {title} {\enquote {\bibinfo {title} {{A membrane paradigm at
  large D}},}\ }\href {\doibase 10.1007/JHEP04(2016)076} {\bibfield  {journal}
  {\bibinfo  {journal} {JHEP}\ }\textbf {\bibinfo {volume} {04}},\ \bibinfo
  {pages} {076} (\bibinfo {year} {2016}{\natexlab{a}})},\ \Eprint
  {http://arxiv.org/abs/1504.06613} {arXiv:1504.06613 [hep-th]} \BibitemShut
  {NoStop}%
\bibitem [{\citenamefont {Bhattacharyya}\ \emph
  {et~al.}(2016{\natexlab{b}})\citenamefont {Bhattacharyya}, \citenamefont
  {Mandlik}, \citenamefont {Minwalla},\ and\ \citenamefont
  {Thakur}}]{Bhattacharyya:2015fdk}%
  \BibitemOpen
  \bibfield  {author} {\bibinfo {author} {\bibfnamefont {Sayantani}\
  \bibnamefont {Bhattacharyya}}, \bibinfo {author} {\bibfnamefont {Mangesh}\
  \bibnamefont {Mandlik}}, \bibinfo {author} {\bibfnamefont {Shiraz}\
  \bibnamefont {Minwalla}}, \ and\ \bibinfo {author} {\bibfnamefont {Somyadip}\
  \bibnamefont {Thakur}},\ }\bibfield  {title} {\enquote {\bibinfo {title} {{A
  Charged Membrane Paradigm at Large D}},}\ }\href {\doibase
  10.1007/JHEP04(2016)128} {\bibfield  {journal} {\bibinfo  {journal} {JHEP}\
  }\textbf {\bibinfo {volume} {04}},\ \bibinfo {pages} {128} (\bibinfo {year}
  {2016}{\natexlab{b}})},\ \Eprint {http://arxiv.org/abs/1511.03432}
  {arXiv:1511.03432 [hep-th]} \BibitemShut {NoStop}%
\bibitem [{\citenamefont {Emparan}\ and\ \citenamefont
  {Herzog}(2020)}]{Emparan:2020inr}%
  \BibitemOpen
  \bibfield  {author} {\bibinfo {author} {\bibfnamefont {Roberto}\ \bibnamefont
  {Emparan}}\ and\ \bibinfo {author} {\bibfnamefont {Christopher~P.}\
  \bibnamefont {Herzog}},\ }\bibfield  {title} {\enquote {\bibinfo {title}
  {{Large D limit of Einstein{\textquoteright}s equations}},}\ }\href {\doibase
  10.1103/RevModPhys.92.045005} {\bibfield  {journal} {\bibinfo  {journal}
  {Rev. Mod. Phys.}\ }\textbf {\bibinfo {volume} {92}},\ \bibinfo {pages}
  {045005} (\bibinfo {year} {2020})},\ \Eprint
  {http://arxiv.org/abs/2003.11394} {arXiv:2003.11394 [hep-th]} \BibitemShut
  {NoStop}%
\bibitem [{\citenamefont {Emparan}\ \emph {et~al.}(2026)\citenamefont
  {Emparan}, \citenamefont {Rafecas-Ventosa},\ and\ \citenamefont
  {Way}}]{Emparan:2025yfy}%
  \BibitemOpen
  \bibfield  {author} {\bibinfo {author} {\bibfnamefont {Roberto}\ \bibnamefont
  {Emparan}}, \bibinfo {author} {\bibfnamefont {Jordi}\ \bibnamefont
  {Rafecas-Ventosa}}, \ and\ \bibinfo {author} {\bibfnamefont {Benson}\
  \bibnamefont {Way}},\ }\bibfield  {title} {\enquote {\bibinfo {title}
  {{General effective theories of black holes in the large D limit}},}\ }\href
  {\doibase 10.1007/JHEP04(2026)034} {\bibfield  {journal} {\bibinfo  {journal}
  {JHEP}\ }\textbf {\bibinfo {volume} {04}},\ \bibinfo {pages} {034} (\bibinfo
  {year} {2026})},\ \Eprint {http://arxiv.org/abs/2512.14186} {arXiv:2512.14186
  [hep-th]} \BibitemShut {NoStop}%
\bibitem [{\citenamefont {Emparan}\ \emph
  {et~al.}(2015{\natexlab{b}})\citenamefont {Emparan}, \citenamefont {Suzuki},\
  and\ \citenamefont {Tanabe}}]{Emparan:2015gva}%
  \BibitemOpen
  \bibfield  {author} {\bibinfo {author} {\bibfnamefont {Roberto}\ \bibnamefont
  {Emparan}}, \bibinfo {author} {\bibfnamefont {Ryotaku}\ \bibnamefont
  {Suzuki}}, \ and\ \bibinfo {author} {\bibfnamefont {Kentaro}\ \bibnamefont
  {Tanabe}},\ }\bibfield  {title} {\enquote {\bibinfo {title} {{Evolution and
  End Point of the Black String Instability: Large D Solution}},}\ }\href
  {\doibase 10.1103/PhysRevLett.115.091102} {\bibfield  {journal} {\bibinfo
  {journal} {Phys. Rev. Lett.}\ }\textbf {\bibinfo {volume} {115}},\ \bibinfo
  {pages} {091102} (\bibinfo {year} {2015}{\natexlab{b}})},\ \Eprint
  {http://arxiv.org/abs/1506.06772} {arXiv:1506.06772 [hep-th]} \BibitemShut
  {NoStop}%
\bibitem [{\citenamefont {Bhattacharyya}\ \emph {et~al.}(2017)\citenamefont
  {Bhattacharyya}, \citenamefont {Mandal}, \citenamefont {Mandlik},
  \citenamefont {Mehta}, \citenamefont {Minwalla}, \citenamefont {Sharma},\
  and\ \citenamefont {Thakur}}]{Bhattacharyya:2016nhn}%
  \BibitemOpen
  \bibfield  {author} {\bibinfo {author} {\bibfnamefont {Sayantani}\
  \bibnamefont {Bhattacharyya}}, \bibinfo {author} {\bibfnamefont {Anup~Kumar}\
  \bibnamefont {Mandal}}, \bibinfo {author} {\bibfnamefont {Mangesh}\
  \bibnamefont {Mandlik}}, \bibinfo {author} {\bibfnamefont {Umang}\
  \bibnamefont {Mehta}}, \bibinfo {author} {\bibfnamefont {Shiraz}\
  \bibnamefont {Minwalla}}, \bibinfo {author} {\bibfnamefont {Utkarsh}\
  \bibnamefont {Sharma}}, \ and\ \bibinfo {author} {\bibfnamefont {Somyadip}\
  \bibnamefont {Thakur}},\ }\bibfield  {title} {\enquote {\bibinfo {title}
  {{Currents and Radiation from the large $D$ Black Hole Membrane}},}\ }\href
  {\doibase 10.1007/JHEP05(2017)098} {\bibfield  {journal} {\bibinfo  {journal}
  {JHEP}\ }\textbf {\bibinfo {volume} {05}},\ \bibinfo {pages} {098} (\bibinfo
  {year} {2017})},\ \Eprint {http://arxiv.org/abs/1611.09310} {arXiv:1611.09310
  [hep-th]} \BibitemShut {NoStop}%
\bibitem [{\citenamefont {Emparan}\ \emph {et~al.}(2014)\citenamefont
  {Emparan}, \citenamefont {Suzuki},\ and\ \citenamefont
  {Tanabe}}]{Emparan:2014aba}%
  \BibitemOpen
  \bibfield  {author} {\bibinfo {author} {\bibfnamefont {Roberto}\ \bibnamefont
  {Emparan}}, \bibinfo {author} {\bibfnamefont {Ryotaku}\ \bibnamefont
  {Suzuki}}, \ and\ \bibinfo {author} {\bibfnamefont {Kentaro}\ \bibnamefont
  {Tanabe}},\ }\bibfield  {title} {\enquote {\bibinfo {title} {{Decoupling and
  non-decoupling dynamics of large D black holes}},}\ }\href {\doibase
  10.1007/JHEP07(2014)113} {\bibfield  {journal} {\bibinfo  {journal} {JHEP}\
  }\textbf {\bibinfo {volume} {07}},\ \bibinfo {pages} {113} (\bibinfo {year}
  {2014})},\ \Eprint {http://arxiv.org/abs/1406.1258} {arXiv:1406.1258
  [hep-th]} \BibitemShut {NoStop}%
\bibitem [{\citenamefont {Emparan}\ \emph
  {et~al.}(2015{\natexlab{c}})\citenamefont {Emparan}, \citenamefont {Suzuki},\
  and\ \citenamefont {Tanabe}}]{Emparan:2015rva}%
  \BibitemOpen
  \bibfield  {author} {\bibinfo {author} {\bibfnamefont {Roberto}\ \bibnamefont
  {Emparan}}, \bibinfo {author} {\bibfnamefont {Ryotaku}\ \bibnamefont
  {Suzuki}}, \ and\ \bibinfo {author} {\bibfnamefont {Kentaro}\ \bibnamefont
  {Tanabe}},\ }\bibfield  {title} {\enquote {\bibinfo {title} {{Quasinormal
  modes of (Anti-)de Sitter black holes in the 1/D expansion}},}\ }\href
  {\doibase 10.1007/JHEP04(2015)085} {\bibfield  {journal} {\bibinfo  {journal}
  {JHEP}\ }\textbf {\bibinfo {volume} {04}},\ \bibinfo {pages} {085} (\bibinfo
  {year} {2015}{\natexlab{c}})},\ \Eprint {http://arxiv.org/abs/1502.02820}
  {arXiv:1502.02820 [hep-th]} \BibitemShut {NoStop}%
\bibitem [{\citenamefont {Emparan}\ \emph {et~al.}(2018)\citenamefont
  {Emparan}, \citenamefont {Luna}, \citenamefont {Mart{\'\i}nez}, \citenamefont
  {Suzuki},\ and\ \citenamefont {Tanabe}}]{Emparan:2018bmi}%
  \BibitemOpen
  \bibfield  {author} {\bibinfo {author} {\bibfnamefont {Roberto}\ \bibnamefont
  {Emparan}}, \bibinfo {author} {\bibfnamefont {Raimon}\ \bibnamefont {Luna}},
  \bibinfo {author} {\bibfnamefont {Marina}\ \bibnamefont {Mart{\'\i}nez}},
  \bibinfo {author} {\bibfnamefont {Ryotaku}\ \bibnamefont {Suzuki}}, \ and\
  \bibinfo {author} {\bibfnamefont {Kentaro}\ \bibnamefont {Tanabe}},\
  }\bibfield  {title} {\enquote {\bibinfo {title} {{Phases and Stability of
  Non-Uniform Black Strings}},}\ }\href {\doibase 10.1007/JHEP05(2018)104}
  {\bibfield  {journal} {\bibinfo  {journal} {JHEP}\ }\textbf {\bibinfo
  {volume} {05}},\ \bibinfo {pages} {104} (\bibinfo {year} {2018})},\ \Eprint
  {http://arxiv.org/abs/1802.08191} {arXiv:1802.08191 [hep-th]} \BibitemShut
  {NoStop}%
\bibitem [{\citenamefont {Andrade}\ \emph
  {et~al.}(2019{\natexlab{a}})\citenamefont {Andrade}, \citenamefont {Emparan},
  \citenamefont {Licht},\ and\ \citenamefont {Luna}}]{Andrade:2019edf}%
  \BibitemOpen
  \bibfield  {author} {\bibinfo {author} {\bibfnamefont {Tom{\'a}s}\
  \bibnamefont {Andrade}}, \bibinfo {author} {\bibfnamefont {Roberto}\
  \bibnamefont {Emparan}}, \bibinfo {author} {\bibfnamefont {David}\
  \bibnamefont {Licht}}, \ and\ \bibinfo {author} {\bibfnamefont {Raimon}\
  \bibnamefont {Luna}},\ }\bibfield  {title} {\enquote {\bibinfo {title}
  {{Black hole collisions, instabilities, and cosmic censorship violation at
  large $D$}},}\ }\href {\doibase 10.1007/JHEP09(2019)099} {\bibfield
  {journal} {\bibinfo  {journal} {JHEP}\ }\textbf {\bibinfo {volume} {09}},\
  \bibinfo {pages} {099} (\bibinfo {year} {2019}{\natexlab{a}})},\ \Eprint
  {http://arxiv.org/abs/1908.03424} {arXiv:1908.03424 [hep-th]} \BibitemShut
  {NoStop}%
\bibitem [{\citenamefont {Andrade}\ \emph {et~al.}(2022)\citenamefont
  {Andrade}, \citenamefont {Figueras},\ and\ \citenamefont
  {Sperhake}}]{Andrade:2020dgc}%
  \BibitemOpen
  \bibfield  {author} {\bibinfo {author} {\bibfnamefont {Tomas}\ \bibnamefont
  {Andrade}}, \bibinfo {author} {\bibfnamefont {Pau}\ \bibnamefont {Figueras}},
  \ and\ \bibinfo {author} {\bibfnamefont {Ulrich}\ \bibnamefont {Sperhake}},\
  }\bibfield  {title} {\enquote {\bibinfo {title} {{Evidence for violations of
  Weak Cosmic Censorship in black hole collisions in higher dimensions}},}\
  }\href {\doibase 10.1007/JHEP03(2022)111} {\bibfield  {journal} {\bibinfo
  {journal} {JHEP}\ }\textbf {\bibinfo {volume} {03}},\ \bibinfo {pages} {111}
  (\bibinfo {year} {2022})},\ \Eprint {http://arxiv.org/abs/2011.03049}
  {arXiv:2011.03049 [hep-th]} \BibitemShut {NoStop}%
\bibitem [{\citenamefont {Emparan}\ \emph {et~al.}(2023)\citenamefont
  {Emparan}, \citenamefont {Luna}, \citenamefont {Suzuki}, \citenamefont
  {Toma{\v{s}}evi{\'c}},\ and\ \citenamefont {Way}}]{Emparan:2023dxm}%
  \BibitemOpen
  \bibfield  {author} {\bibinfo {author} {\bibfnamefont {Roberto}\ \bibnamefont
  {Emparan}}, \bibinfo {author} {\bibfnamefont {Raimon}\ \bibnamefont {Luna}},
  \bibinfo {author} {\bibfnamefont {Ryotaku}\ \bibnamefont {Suzuki}}, \bibinfo
  {author} {\bibfnamefont {Marija}\ \bibnamefont {Toma{\v{s}}evi{\'c}}}, \ and\
  \bibinfo {author} {\bibfnamefont {Benson}\ \bibnamefont {Way}},\ }\bibfield
  {title} {\enquote {\bibinfo {title} {{Holographic duals of evaporating black
  holes}},}\ }\href {\doibase 10.1007/JHEP05(2023)182} {\bibfield  {journal}
  {\bibinfo  {journal} {JHEP}\ }\textbf {\bibinfo {volume} {05}},\ \bibinfo
  {pages} {182} (\bibinfo {year} {2023})},\ \Eprint
  {http://arxiv.org/abs/2301.02587} {arXiv:2301.02587 [hep-th]} \BibitemShut
  {NoStop}%
\bibitem [{\citenamefont {Emparan}\ \emph {et~al.}(2016)\citenamefont
  {Emparan}, \citenamefont {Izumi}, \citenamefont {Luna}, \citenamefont
  {Suzuki},\ and\ \citenamefont {Tanabe}}]{Emparan:2016sjk}%
  \BibitemOpen
  \bibfield  {author} {\bibinfo {author} {\bibfnamefont {Roberto}\ \bibnamefont
  {Emparan}}, \bibinfo {author} {\bibfnamefont {Keisuke}\ \bibnamefont
  {Izumi}}, \bibinfo {author} {\bibfnamefont {Raimon}\ \bibnamefont {Luna}},
  \bibinfo {author} {\bibfnamefont {Ryotaku}\ \bibnamefont {Suzuki}}, \ and\
  \bibinfo {author} {\bibfnamefont {Kentaro}\ \bibnamefont {Tanabe}},\
  }\bibfield  {title} {\enquote {\bibinfo {title} {{Hydro-elastic
  Complementarity in Black Branes at large D}},}\ }\href {\doibase
  10.1007/JHEP06(2016)117} {\bibfield  {journal} {\bibinfo  {journal} {JHEP}\
  }\textbf {\bibinfo {volume} {06}},\ \bibinfo {pages} {117} (\bibinfo {year}
  {2016})},\ \Eprint {http://arxiv.org/abs/1602.05752} {arXiv:1602.05752
  [hep-th]} \BibitemShut {NoStop}%
\bibitem [{\citenamefont {Andrade}\ \emph {et~al.}(2018)\citenamefont
  {Andrade}, \citenamefont {Emparan},\ and\ \citenamefont
  {Licht}}]{Andrade:2018nsz}%
  \BibitemOpen
  \bibfield  {author} {\bibinfo {author} {\bibfnamefont {Tom{\'a}s}\
  \bibnamefont {Andrade}}, \bibinfo {author} {\bibfnamefont {Roberto}\
  \bibnamefont {Emparan}}, \ and\ \bibinfo {author} {\bibfnamefont {David}\
  \bibnamefont {Licht}},\ }\bibfield  {title} {\enquote {\bibinfo {title}
  {{Rotating black holes and black bars at large D}},}\ }\href {\doibase
  10.1007/JHEP09(2018)107} {\bibfield  {journal} {\bibinfo  {journal} {JHEP}\
  }\textbf {\bibinfo {volume} {09}},\ \bibinfo {pages} {107} (\bibinfo {year}
  {2018})},\ \Eprint {http://arxiv.org/abs/1807.01131} {arXiv:1807.01131
  [hep-th]} \BibitemShut {NoStop}%
\bibitem [{\citenamefont {Andrade}\ \emph
  {et~al.}(2019{\natexlab{b}})\citenamefont {Andrade}, \citenamefont {Emparan},
  \citenamefont {Licht},\ and\ \citenamefont {Luna}}]{Andrade:2018yqu}%
  \BibitemOpen
  \bibfield  {author} {\bibinfo {author} {\bibfnamefont {Tom{\'a}s}\
  \bibnamefont {Andrade}}, \bibinfo {author} {\bibfnamefont {Roberto}\
  \bibnamefont {Emparan}}, \bibinfo {author} {\bibfnamefont {David}\
  \bibnamefont {Licht}}, \ and\ \bibinfo {author} {\bibfnamefont {Raimon}\
  \bibnamefont {Luna}},\ }\bibfield  {title} {\enquote {\bibinfo {title}
  {{Cosmic censorship violation in black hole collisions in higher
  dimensions}},}\ }\href {\doibase 10.1007/JHEP04(2019)121} {\bibfield
  {journal} {\bibinfo  {journal} {JHEP}\ }\textbf {\bibinfo {volume} {04}},\
  \bibinfo {pages} {121} (\bibinfo {year} {2019}{\natexlab{b}})},\ \Eprint
  {http://arxiv.org/abs/1812.05017} {arXiv:1812.05017 [hep-th]} \BibitemShut
  {NoStop}%
\bibitem [{\citenamefont {Andrade}\ \emph {et~al.}(2020)\citenamefont
  {Andrade}, \citenamefont {Emparan}, \citenamefont {Jansen}, \citenamefont
  {Licht}, \citenamefont {Luna},\ and\ \citenamefont
  {Suzuki}}]{Andrade:2020ilm}%
  \BibitemOpen
  \bibfield  {author} {\bibinfo {author} {\bibfnamefont {Tomas}\ \bibnamefont
  {Andrade}}, \bibinfo {author} {\bibfnamefont {Roberto}\ \bibnamefont
  {Emparan}}, \bibinfo {author} {\bibfnamefont {Aron}\ \bibnamefont {Jansen}},
  \bibinfo {author} {\bibfnamefont {David}\ \bibnamefont {Licht}}, \bibinfo
  {author} {\bibfnamefont {Raimon}\ \bibnamefont {Luna}}, \ and\ \bibinfo
  {author} {\bibfnamefont {Ryotaku}\ \bibnamefont {Suzuki}},\ }\bibfield
  {title} {\enquote {\bibinfo {title} {{Entropy production and entropic
  attractors in black hole fusion and fission}},}\ }\href {\doibase
  10.1007/JHEP08(2020)098} {\bibfield  {journal} {\bibinfo  {journal} {JHEP}\
  }\textbf {\bibinfo {volume} {08}},\ \bibinfo {pages} {098} (\bibinfo {year}
  {2020})},\ \Eprint {http://arxiv.org/abs/2005.14498} {arXiv:2005.14498
  [hep-th]} \BibitemShut {NoStop}%
\bibitem [{\citenamefont {Leaver}(1986)}]{Leaver_prop}%
  \BibitemOpen
  \bibfield  {author} {\bibinfo {author} {\bibfnamefont {Edward~W.}\
  \bibnamefont {Leaver}},\ }\bibfield  {title} {\enquote {\bibinfo {title}
  {Spectral decomposition of the perturbation response of the schwarzschild
  geometry},}\ }\href {\doibase 10.1103/PhysRevD.34.384} {\bibfield  {journal}
  {\bibinfo  {journal} {Phys. Rev. D}\ }\textbf {\bibinfo {volume} {34}},\
  \bibinfo {pages} {384--408} (\bibinfo {year} {1986})}\BibitemShut {NoStop}%
\bibitem [{\citenamefont {Fransen}\ \emph {et~al.}(2025)\citenamefont
  {Fransen}, \citenamefont {Pere{\~n}iguez},\ and\ \citenamefont
  {Redondo-Yuste}}]{Fransen:2025cgv}%
  \BibitemOpen
  \bibfield  {author} {\bibinfo {author} {\bibfnamefont {Kwinten}\ \bibnamefont
  {Fransen}}, \bibinfo {author} {\bibfnamefont {David}\ \bibnamefont
  {Pere{\~n}iguez}}, \ and\ \bibinfo {author} {\bibfnamefont {Jaime}\
  \bibnamefont {Redondo-Yuste}},\ }\bibfield  {title} {\enquote {\bibinfo
  {title} {{Perturbations of plane waves and quadratic quasinormal modes on the
  lightring}},}\ }\href {\doibase 10.1007/JHEP12(2025)148} {\bibfield
  {journal} {\bibinfo  {journal} {JHEP}\ }\textbf {\bibinfo {volume} {12}},\
  \bibinfo {pages} {148} (\bibinfo {year} {2025})},\ \Eprint
  {http://arxiv.org/abs/2509.03598} {arXiv:2509.03598 [gr-qc]} \BibitemShut
  {NoStop}%
\bibitem [{\citenamefont {London}\ and\ \citenamefont
  {Foucoin}(2026)}]{London:2023idh}%
  \BibitemOpen
  \bibfield  {author} {\bibinfo {author} {\bibfnamefont {Lionel}\ \bibnamefont
  {London}}\ and\ \bibinfo {author} {\bibfnamefont {Michelle}\ \bibnamefont
  {Foucoin}},\ }\bibfield  {title} {\enquote {\bibinfo {title} {{Natural
  polynomials for Kerr quasinormal modes}},}\ }\href {\doibase
  10.1103/nn8t-p14d} {\bibfield  {journal} {\bibinfo  {journal} {Phys. Rev. D}\
  }\textbf {\bibinfo {volume} {113}},\ \bibinfo {pages} {044009} (\bibinfo
  {year} {2026})},\ \Eprint {http://arxiv.org/abs/2312.17680} {arXiv:2312.17680
  [gr-qc]} \BibitemShut {NoStop}%
\bibitem [{\citenamefont {Dyer}\ and\ \citenamefont
  {Moore}(2025)}]{Dyer:2025hdt}%
  \BibitemOpen
  \bibfield  {author} {\bibinfo {author} {\bibfnamefont {Richard}\ \bibnamefont
  {Dyer}}\ and\ \bibinfo {author} {\bibfnamefont {Christopher~J.}\ \bibnamefont
  {Moore}},\ }\bibfield  {title} {\enquote {\bibinfo {title} {{The quasinormal
  mode content of binary black hole ringdown}},}\ }\href@noop {} {\  (\bibinfo
  {year} {2025})},\ \Eprint {http://arxiv.org/abs/2510.13954} {arXiv:2510.13954
  [gr-qc]} \BibitemShut {NoStop}%
\bibitem [{\citenamefont {Berti}\ \emph {et~al.}(2004)\citenamefont {Berti},
  \citenamefont {Cavaglia},\ and\ \citenamefont {Gualtieri}}]{Berti:2003si}%
  \BibitemOpen
  \bibfield  {author} {\bibinfo {author} {\bibfnamefont {Emanuele}\
  \bibnamefont {Berti}}, \bibinfo {author} {\bibfnamefont {Marco}\ \bibnamefont
  {Cavaglia}}, \ and\ \bibinfo {author} {\bibfnamefont {Leonardo}\ \bibnamefont
  {Gualtieri}},\ }\bibfield  {title} {\enquote {\bibinfo {title}
  {{Gravitational energy loss in high-energy particle collisions:
  Ultrarelativistic plunge into a multidimensional black hole}},}\ }\href
  {\doibase 10.1103/PhysRevD.69.124011} {\bibfield  {journal} {\bibinfo
  {journal} {Phys. Rev. D}\ }\textbf {\bibinfo {volume} {69}},\ \bibinfo
  {pages} {124011} (\bibinfo {year} {2004})},\ \Eprint
  {http://arxiv.org/abs/hep-th/0309203} {arXiv:hep-th/0309203} \BibitemShut
  {NoStop}%
\bibitem [{\citenamefont {Berti}\ \emph {et~al.}(2011)\citenamefont {Berti},
  \citenamefont {Cardoso},\ and\ \citenamefont {Kipapa}}]{Berti:2010gx}%
  \BibitemOpen
  \bibfield  {author} {\bibinfo {author} {\bibfnamefont {Emanuele}\
  \bibnamefont {Berti}}, \bibinfo {author} {\bibfnamefont {Vitor}\ \bibnamefont
  {Cardoso}}, \ and\ \bibinfo {author} {\bibfnamefont {Barnabas}\ \bibnamefont
  {Kipapa}},\ }\bibfield  {title} {\enquote {\bibinfo {title} {{Up to eleven:
  radiation from particles with arbitrary energy falling into
  higher-dimensional black holes}},}\ }\href {\doibase
  10.1103/PhysRevD.83.084018} {\bibfield  {journal} {\bibinfo  {journal} {Phys.
  Rev. D}\ }\textbf {\bibinfo {volume} {83}},\ \bibinfo {pages} {084018}
  (\bibinfo {year} {2011})},\ \Eprint {http://arxiv.org/abs/1010.3874}
  {arXiv:1010.3874 [gr-qc]} \BibitemShut {NoStop}%
\bibitem [{\citenamefont {Virtanen}\ \emph {et~al.}(2020)\citenamefont
  {Virtanen}, \citenamefont {Gommers}, \citenamefont {Oliphant}, \citenamefont
  {Haberland}, \citenamefont {Reddy}, \citenamefont {Cournapeau}, \citenamefont
  {Burovski}, \citenamefont {Peterson}, \citenamefont {Weckesser},
  \citenamefont {Bright} \emph {et~al.}}]{virtanen2020scipy}%
  \BibitemOpen
  \bibfield  {author} {\bibinfo {author} {\bibfnamefont {Pauli}\ \bibnamefont
  {Virtanen}}, \bibinfo {author} {\bibfnamefont {Ralf}\ \bibnamefont
  {Gommers}}, \bibinfo {author} {\bibfnamefont {Travis~E}\ \bibnamefont
  {Oliphant}}, \bibinfo {author} {\bibfnamefont {Matt}\ \bibnamefont
  {Haberland}}, \bibinfo {author} {\bibfnamefont {Tyler}\ \bibnamefont
  {Reddy}}, \bibinfo {author} {\bibfnamefont {David}\ \bibnamefont
  {Cournapeau}}, \bibinfo {author} {\bibfnamefont {Evgeni}\ \bibnamefont
  {Burovski}}, \bibinfo {author} {\bibfnamefont {Pearu}\ \bibnamefont
  {Peterson}}, \bibinfo {author} {\bibfnamefont {Warren}\ \bibnamefont
  {Weckesser}}, \bibinfo {author} {\bibfnamefont {Jonathan}\ \bibnamefont
  {Bright}},  \emph {et~al.},\ }\bibfield  {title} {\enquote {\bibinfo {title}
  {Scipy 1.0: fundamental algorithms for scientific computing in python},}\
  }\href@noop {} {\bibfield  {journal} {\bibinfo  {journal} {Nature methods}\
  }\textbf {\bibinfo {volume} {17}},\ \bibinfo {pages} {261--272} (\bibinfo
  {year} {2020})}\BibitemShut {NoStop}%
\bibitem [{\citenamefont {Baker}\ \emph {et~al.}(2000)\citenamefont {Baker},
  \citenamefont {Bruegmann}, \citenamefont {Campanelli},\ and\ \citenamefont
  {Lousto}}]{Baker:2000zh}%
  \BibitemOpen
  \bibfield  {author} {\bibinfo {author} {\bibfnamefont {John~G.}\ \bibnamefont
  {Baker}}, \bibinfo {author} {\bibfnamefont {Bernd}\ \bibnamefont
  {Bruegmann}}, \bibinfo {author} {\bibfnamefont {Manuela}\ \bibnamefont
  {Campanelli}}, \ and\ \bibinfo {author} {\bibfnamefont {Carlos~O.}\
  \bibnamefont {Lousto}},\ }\bibfield  {title} {\enquote {\bibinfo {title}
  {{Gravitational waves from black hole collisions via an eclectic
  approach}},}\ }\href {\doibase 10.1088/0264-9381/17/20/102} {\bibfield
  {journal} {\bibinfo  {journal} {Class. Quant. Grav.}\ }\textbf {\bibinfo
  {volume} {17}},\ \bibinfo {pages} {L149--L156} (\bibinfo {year} {2000})},\
  \Eprint {http://arxiv.org/abs/gr-qc/0003027} {arXiv:gr-qc/0003027}
  \BibitemShut {NoStop}%
\bibitem [{\citenamefont {Baker}\ \emph {et~al.}(2001)\citenamefont {Baker},
  \citenamefont {Bruegmann}, \citenamefont {Campanelli}, \citenamefont
  {Lousto},\ and\ \citenamefont {Takahashi}}]{Baker:2001nu}%
  \BibitemOpen
  \bibfield  {author} {\bibinfo {author} {\bibfnamefont {John~G.}\ \bibnamefont
  {Baker}}, \bibinfo {author} {\bibfnamefont {Bernd}\ \bibnamefont
  {Bruegmann}}, \bibinfo {author} {\bibfnamefont {Manuela}\ \bibnamefont
  {Campanelli}}, \bibinfo {author} {\bibfnamefont {C.~O.}\ \bibnamefont
  {Lousto}}, \ and\ \bibinfo {author} {\bibfnamefont {R.}~\bibnamefont
  {Takahashi}},\ }\bibfield  {title} {\enquote {\bibinfo {title} {{Plunge wave
  forms from inspiralling binary black holes}},}\ }\href {\doibase
  10.1103/PhysRevLett.87.121103} {\bibfield  {journal} {\bibinfo  {journal}
  {Phys. Rev. Lett.}\ }\textbf {\bibinfo {volume} {87}},\ \bibinfo {pages}
  {121103} (\bibinfo {year} {2001})},\ \Eprint
  {http://arxiv.org/abs/gr-qc/0102037} {arXiv:gr-qc/0102037} \BibitemShut
  {NoStop}%
\bibitem [{\citenamefont {Baker}\ \emph {et~al.}(2002)\citenamefont {Baker},
  \citenamefont {Campanelli},\ and\ \citenamefont {Lousto}}]{Baker:2001sf}%
  \BibitemOpen
  \bibfield  {author} {\bibinfo {author} {\bibfnamefont {John~G.}\ \bibnamefont
  {Baker}}, \bibinfo {author} {\bibfnamefont {Manuela}\ \bibnamefont
  {Campanelli}}, \ and\ \bibinfo {author} {\bibfnamefont {Carlos~O.}\
  \bibnamefont {Lousto}},\ }\bibfield  {title} {\enquote {\bibinfo {title}
  {{The Lazarus project: A Pragmatic approach to binary black hole
  evolutions}},}\ }\href {\doibase 10.1103/PhysRevD.65.044001} {\bibfield
  {journal} {\bibinfo  {journal} {Phys. Rev. D}\ }\textbf {\bibinfo {volume}
  {65}},\ \bibinfo {pages} {044001} (\bibinfo {year} {2002})},\ \Eprint
  {http://arxiv.org/abs/gr-qc/0104063} {arXiv:gr-qc/0104063} \BibitemShut
  {NoStop}%
\bibitem [{\citenamefont {Campanelli}\ \emph {et~al.}(2006)\citenamefont
  {Campanelli}, \citenamefont {Kelly},\ and\ \citenamefont
  {Lousto}}]{Campanelli:2005ia}%
  \BibitemOpen
  \bibfield  {author} {\bibinfo {author} {\bibfnamefont {Manuela}\ \bibnamefont
  {Campanelli}}, \bibinfo {author} {\bibfnamefont {Bernard~J.}\ \bibnamefont
  {Kelly}}, \ and\ \bibinfo {author} {\bibfnamefont {Carlos~O.}\ \bibnamefont
  {Lousto}},\ }\bibfield  {title} {\enquote {\bibinfo {title} {{The Lazarus
  project. II. Space-like extraction with the quasi-Kinnersley tetrad}},}\
  }\href {\doibase 10.1103/PhysRevD.73.064005} {\bibfield  {journal} {\bibinfo
  {journal} {Phys. Rev. D}\ }\textbf {\bibinfo {volume} {73}},\ \bibinfo
  {pages} {064005} (\bibinfo {year} {2006})},\ \Eprint
  {http://arxiv.org/abs/gr-qc/0510122} {arXiv:gr-qc/0510122} \BibitemShut
  {NoStop}%
\bibitem [{\citenamefont {Price}\ and\ \citenamefont
  {Pullin}(1994)}]{Price:1994pm}%
  \BibitemOpen
  \bibfield  {author} {\bibinfo {author} {\bibfnamefont {Richard~H.}\
  \bibnamefont {Price}}\ and\ \bibinfo {author} {\bibfnamefont {Jorge}\
  \bibnamefont {Pullin}},\ }\bibfield  {title} {\enquote {\bibinfo {title}
  {{Colliding black holes: The Close limit}},}\ }\href {\doibase
  10.1103/PhysRevLett.72.3297} {\bibfield  {journal} {\bibinfo  {journal}
  {Phys. Rev. Lett.}\ }\textbf {\bibinfo {volume} {72}},\ \bibinfo {pages}
  {3297--3300} (\bibinfo {year} {1994})},\ \Eprint
  {http://arxiv.org/abs/gr-qc/9402039} {arXiv:gr-qc/9402039} \BibitemShut
  {NoStop}%
\bibitem [{\citenamefont {Khanna}\ \emph {et~al.}(1999)\citenamefont {Khanna},
  \citenamefont {Baker}, \citenamefont {Gleiser}, \citenamefont {Laguna},
  \citenamefont {Nicasio}, \citenamefont {Nollert}, \citenamefont {Price},\
  and\ \citenamefont {Pullin}}]{Khanna:1999mh}%
  \BibitemOpen
  \bibfield  {author} {\bibinfo {author} {\bibfnamefont {Gaurav}\ \bibnamefont
  {Khanna}}, \bibinfo {author} {\bibfnamefont {John~G.}\ \bibnamefont {Baker}},
  \bibinfo {author} {\bibfnamefont {Reinaldo~J.}\ \bibnamefont {Gleiser}},
  \bibinfo {author} {\bibfnamefont {Pablo}\ \bibnamefont {Laguna}}, \bibinfo
  {author} {\bibfnamefont {Carlos~O.}\ \bibnamefont {Nicasio}}, \bibinfo
  {author} {\bibfnamefont {Hans-Peter}\ \bibnamefont {Nollert}}, \bibinfo
  {author} {\bibfnamefont {Richard}\ \bibnamefont {Price}}, \ and\ \bibinfo
  {author} {\bibfnamefont {Jorge}\ \bibnamefont {Pullin}},\ }\bibfield  {title}
  {\enquote {\bibinfo {title} {{Inspiralling black holes: The Close limit}},}\
  }\href {\doibase 10.1103/PhysRevLett.83.3581} {\bibfield  {journal} {\bibinfo
   {journal} {Phys. Rev. Lett.}\ }\textbf {\bibinfo {volume} {83}},\ \bibinfo
  {pages} {3581--3584} (\bibinfo {year} {1999})},\ \Eprint
  {http://arxiv.org/abs/gr-qc/9905081} {arXiv:gr-qc/9905081} \BibitemShut
  {NoStop}%
\bibitem [{\citenamefont {Ma}\ \emph {et~al.}(2024)\citenamefont {Ma} \emph
  {et~al.}}]{Ma:2023qjn}%
  \BibitemOpen
  \bibfield  {author} {\bibinfo {author} {\bibfnamefont {Sizheng}\ \bibnamefont
  {Ma}} \emph {et~al.},\ }\bibfield  {title} {\enquote {\bibinfo {title}
  {{Fully relativistic three-dimensional Cauchy-characteristic matching for
  physical degrees of freedom}},}\ }\href {\doibase
  10.1103/PhysRevD.109.124027} {\bibfield  {journal} {\bibinfo  {journal}
  {Phys. Rev. D}\ }\textbf {\bibinfo {volume} {109}},\ \bibinfo {pages}
  {124027} (\bibinfo {year} {2024})},\ \Eprint
  {http://arxiv.org/abs/2308.10361} {arXiv:2308.10361 [gr-qc]} \BibitemShut
  {NoStop}%
\bibitem [{\citenamefont {Scheel}\ \emph {et~al.}(2025)\citenamefont {Scheel}
  \emph {et~al.}}]{Scheel:2025jct}%
  \BibitemOpen
  \bibfield  {author} {\bibinfo {author} {\bibfnamefont {Mark~A.}\ \bibnamefont
  {Scheel}} \emph {et~al.},\ }\bibfield  {title} {\enquote {\bibinfo {title}
  {{The SXS collaboration{\textquoteright}s third catalog of binary black hole
  simulations}},}\ }\href {\doibase 10.1088/1361-6382/adfd34} {\bibfield
  {journal} {\bibinfo  {journal} {Class. Quant. Grav.}\ }\textbf {\bibinfo
  {volume} {42}},\ \bibinfo {pages} {195017} (\bibinfo {year} {2025})},\
  \Eprint {http://arxiv.org/abs/2505.13378} {arXiv:2505.13378 [gr-qc]}
  \BibitemShut {NoStop}%
\end{thebibliography}%

\appendix
\onecolumngrid
\section{Quadratic solution from $\ell=2$ and $\ell=3$ linear modes}\label{ap:n2n3}

At second order, we find four oscillatory frequency channels
\begin{equation}
    \begin{aligned}
        &\sigma_{2,1}=2-2i\, ,\quad  \sigma_{2,2}=2\sqrt{2}-4i\, ,\\
        &\sigma_{2,3}=1+\sqrt{2}-3i\, ,\quad \sigma_{2,4}=1-\sqrt{2}-3i\, .
    \end{aligned}
\end{equation}
and two purely damped ones,
\begin{equation}
   \sigma_{2,\circ1}=-2i\, ,\qquad \sigma_{2,\circ2}=-4i\, .
\end{equation}
One can also read certain parity-selection rules simply from the parity properties of Hermite polynomials. In particular, for configurations that are symmetric under $z\to -z$ all odd $\ell$ should vanish, and this symmetry is obviously preserved beyond linear order.
The channels $\sigma_{2,1},\sigma_{2,\circ 1}$ and $\sigma_{2,2},\sigma_{2,\circ 2}$ come from the self coupling of the $\ell=2$ and $\ell=3$ modes with themselves, and $\sigma_{2,3},\sigma_{2,4}$ arise from their mutual interaction.

The amplitudes of the oscillatory channels are
    \begin{equation}\label{eq:A_osc}
\begin{aligned}
    \sigma_{2,1}:\qquad &A^{2}_{2,1}=A_{2}^2\left(-\frac{1}{5}+\frac{i}{10}\right)  e^{-2 i \phi_{2}}\, , \ \ \ A^{4}_{2,1}= A_{2}^2\frac{1}{8} e^{-2 i \phi_{2}}\, ,\\ \\
    \sigma_{2,2}:\qquad &A^{2}_{2,2}= A_{3}^2\frac{1}{73} \left(-25+4 i \sqrt{2}\right) e^{-2 i \phi_{3}}\,,\ \ \ A^{4}_{2,2}= A_{3}^2\frac{1}{2} i \left(\sqrt{2}+i\right) e^{-2 i \phi_{3}}\,,\ \ \ A^{6}_{2,2}=A_{3}^2\frac{1}{8}  e^{-2 i \phi_{3}}\, ,\\ \\
    \sigma_{2,3}:\qquad &A^{3}_{2,3}=A_{2} A_{3}\frac{(1+i) \left((1+2 i)-(1-2 i) \sqrt{2}\right)  e^{-i (\phi_{2}+\phi_{3})}}{2 \sqrt{2}+5}\,,\ \ \  A^{5}_{2,3}=A_{2} A_{3}\frac{1}{4}  e^{-i (\phi_{2}+\phi_{3})}\,,\\ \\ 
     \sigma_{2,4}:\qquad &A^{3}_{2,4}= A_{2} A_{3}\frac{\left((1-3 i)-(3-i) \sqrt{2}\right) e^{-i (\phi_{2}-\phi_{3})}}{2 \sqrt{2}-5}\,,\ \ \  A^{5}_{2,4}= A_{2} A_{3} \frac{1}{4}  e^{-i (\phi_{2}-\phi_{3})}\, ,
\end{aligned}
\end{equation}
while those of the purely damped ones are
\begin{equation}\label{eq:A_damp}
\begin{aligned}
    \sigma_{2,\circ 1}:\qquad &A^{2}_{2,\circ1}=A^{2}_{2}\, ,\ \ \ A^{4}_{2,\circ1}=A^{2}_{2}\frac{1}{4}\, ,\\ \\
    \sigma_{2,\circ 2}:\qquad &A^{2}_{2,\circ2}=A^{2}_{3}\frac{6}{5}\, ,\ \ \ A^{4}_{2,\circ2}=3 A^{2}_{3}\,, \\ &A^{6}_{2,\circ2}=A^{2}_{3}\frac{1}{4}\,.
\end{aligned}
\end{equation}
The mass coefficients $\mathcal{R}^{\ell}_{2}(t)$ become
\begin{equation}
    \begin{aligned}
        \mathcal{R}^{2}_{2}(t)&=-\frac{1}{365} e^{-4 t} \Bigl(-73 A_{2}^2 e^{2 t} \sin (2 t+2 \phi_{2})+146 A_{2}^2 e^{2 t} \cos (2 t+2 \phi_{2})-365 A_{2}^2 e^{2 t}-40 \sqrt{2} A_{3}^2 \sin \left(2 \sqrt{2} t+2 \phi_{3}\right)\\
        &+250 A_{3}^2 \cos \left(2 \sqrt{2} t+2 \phi_{3}\right)-438 A_{3}^2\Bigr)\, , \\ \\ 
        \mathcal{R}^{3}_{2}(t)&=\frac{2}{17} \left(13 \sqrt{2}+7\right) A_{2} A_{3} e^{-3 t} \cos \left(-\sqrt{2} t+t+\phi_{2}-\phi_{3}\right)+\frac{2}{17} \left(7-13 \sqrt{2}\right) A_{2} A_{3} e^{-3 t} \cos \left(\sqrt{2} t+t+\phi_{2}+\phi_{3}\right)\\
        &+\frac{2}{17} \left(11-\sqrt{2}\right) A_{2} A_{3} e^{-3 t} \sin \left(\sqrt{2} t+t+\phi_{2}+\phi_{3}\right)+\frac{2}{17} \left(\sqrt{2}+11\right) A_{2} A_{3} e^{-3 t} \sin \left(-\sqrt{2} t+t+\phi_{2}-\phi_{3}\right)\, ,\\ \\ 
        \mathcal{R}^{4}_{2}(t)&= \frac{1}{4} e^{-4 t} \left(A_{2}^2 e^{2 t} \cos (2 t+2 \phi_{2})+A_{2}^2 e^{2 t}+4 \sqrt{2} A_{3}^2 \sin \left(2 \sqrt{2} t+2 \phi_{3}\right)-4 A_{3}^2 \cos \left(2 \sqrt{2} t+2 \phi_{3}\right)+12 A_{3}^2\right)\, ,\\ \\ 
        \mathcal{R}^{5}_{2}(t)&=\frac{1}{2} A_{2} A_{3} e^{-3 t} \cos \left(-\sqrt{2} t+t+\phi_{2}-\phi_{3}\right)+\frac{1}{2} A_{2} A_{3} e^{-3 t} \cos \left(\sqrt{2} t+t+\phi_{2}+\phi_{3}\right)\, , \\ \\ 
        \mathcal{R}^{6}_{2}(t)&= \frac{1}{4} A_{3}^2 e^{-4 t} \left(\cos \left(2 \sqrt{2} t+2 \phi_{3}\right)+1\right)\, . 
    \end{aligned}
\end{equation}
\end{document}